\documentclass[aps, prl, 10pt, twocolumn,superscriptaddress]{revtex4-1}

\usepackage[utf8]{inputenc}
\usepackage{epsfig}
\usepackage{graphicx}
\usepackage{float}
\usepackage{dcolumn}
\usepackage{xcolor}
\usepackage{bm}
\usepackage{amsmath,bbm}

\DeclareUnicodeCharacter{2212}{-}

\begin{document}

\title{Ubiquitous non-Majorana Zero-Bias Conductance Peaks in Nanowire Devices}

\author{J. Chen}
\affiliation{Department of Physics and Astronomy, University of Pittsburgh, Pittsburgh, PA 15260, USA} 
\affiliation{Department of Electrical and Computer Engineering and Peterson Institute of NanoScience and Engineering, University of Pittsburgh, Pittsburgh, PA 15261, USA} 
\author{B.D. Woods}
\affiliation{Department of Physics and Astronomy, West Virginia University, Morgantown, WV 26506, USA}    
\author{P. Yu}
\affiliation{Department of Physics and Astronomy, University of Pittsburgh, Pittsburgh, PA 15260, USA} 
\author{M. Hocevar}
\affiliation{Univ. Grenoble Alpes, CNRS, Grenoble INP, Institut N\'eel, 38000 Grenoble, France}
\author{D. Car}
\affiliation{Eindhoven University of Technology, 5600 MB, Eindhoven, The Netherlands} 
\author{S.R. Plissard}
\affiliation{LAAS CNRS, Universit{\'e} de Toulouse, 31031 Toulouse, France} 
\author{E.P.A.M. Bakkers}
\affiliation{Eindhoven University of Technology, 5600 MB, Eindhoven, The Netherlands} 
\author{T.D. Stanescu}
\affiliation{Department of Physics and Astronomy, West Virginia University, Morgantown, WV 26506, USA}  
\author{S.M. Frolov}
\email{frolovsm@pitt.edu}
\affiliation{Department of Physics and Astronomy, University of Pittsburgh, Pittsburgh, PA 15260, USA} 

\date{\today}

\begin{abstract}
We perform tunneling measurements on indium antimonide nanowire/superconductor hybrid devices fabricated for the studies of Majorana bound states. At finite magnetic field, resonances that strongly resemble Majorana bound states, including zero-bias pinning, become common to the point of ubiquity. Since Majorana bound states are predicted in only a limited parameter range in nanowire devices, we seek an alternative explanation for the observed zero-bias peaks. With the help of a  self-consistent Poission-Schr\"odinger multiband model developed in parallel, we identify several families of trivial subgap states which overlap and interact, giving rise to a crowded spectrum near zero energy and zero-bias conductance peaks in experiments. These findings advance the search for Majorana bound states through improved understanding of broader phenomena found in superconductor-semiconductor systems.
\end{abstract}

\maketitle


Majorana bound states (MBS) are predicted in various intrinsic and engineered topological superconductors \cite{Fu2008PRL, AliceaPRB2010,SauPRB2010,LutchynPRL2010,OregPRL2010, AliceaPRP2012, BeenakkerARCMP2013}. They attract sustained attention primarily thanks to the hypothesized non-Abelian rules for the two-MBS exchange \cite{ReadPRB2000}. Tunneling experiments reported signatures of MBS by studying zero-bias conductance peaks   \cite{MourikScience2012, DasNatphys2012, DengNanolett2012, FinckPRL2013, ChurchillPRB2013, Nadj-PergeScience2014,Albrecht2016Nature,  DengScience2016, JChen2017, gulnatnano2018, ZhangNature2018, SuominenPRL2017, NichelePRL2017}. The primary challenge for the tunneling evidence is that zero-bias anomalies in transport are widespread in mesoscopic systems. They have many known non-MBS origins such as Kondo effect \cite{LeePRL2012}, weak antilocalization \cite{PikulinNJP2012}, reflectionless tunneling \cite{PopinciucPRB12}, and supercurrent \cite{ZuoPRL17}. Luckily, most of these phenomena can be ruled out for each particular Majorana experiment through their distinct dependence on the in-situ tunable parameters or through device design. 

Yet, zero-bias anomalies of non-topological origin that closely resemble MBS, and cannot be straightforwardly ruled out, have also been identified. Most remarkably, trivial Andreev Bound States (ABS) have been demonstrated to result in zero-bias peaks \cite{LeeNatnano2014}. This includes peaks that appear at finite magnetic field and exhibit some degree of pinning to zero bias or near-zero oscillations, both being features that MBS and ABS share. Trivial ABS can exist both in the topologically superconducting regime and in the trivial regime, or they can be a result of strong MBS hybridization \cite{KellsPRB2012, Moore2018, vuik2018}. Thus ABS can be found in a much wider range of system parameters than MBS. Understanding of the full ABS phenomenology is therefore central to the unambiguous demonstration of MBS.

In this manuscript, we demonstrate that multiple coexisting and coupled ABS can lead to ubiquitous zero-bias peaks that share spectroscopic signatures with MBS. Our NbTiN/InSb devices have been designed for Majorana experiments, and they yield tunneling resonances that pin near zero source-drain voltage bias at finite external magnetic field, as expected for MBS. However, extended gate voltage sweeps reveal multiple families of states localized near the superconductor. We identify these states as being responsible for the omnipresent zero-bias resonances. The frequency of occurrence of zero-bias features, i.e. their ubiquity, makes it highly unlikely that all of them originate from topologically superconducting segments of the nanowire. A self-consistent multiband model developed in parallel \cite{Woods2019} finds a generic presence of overlapping and coupled trivial ABS for the device geometry used in the experiment. The model identifies that trivial ABS can persist near zero bias due to spectral crowding as well as level repulsion.

The basic MBS theories make a number of simple predictions for the tunneling manifestations of MBS in spin-orbit nanowires \cite{LutchynPRL2010, OregPRL2010}. In long quantum wires MBS should only appear within the topologically superconducting phase described by the  boundary equation: $E_{Z} > \sqrt{\Delta^2 + \mu^2}$, where $E_{Z}= g\mu_{B}B/2$ is the Zeeman energy, with $g$ the effective Land{\'e} $g$-factor, $\mu_{B}$ the Bohr magneton. $\Delta$ is the induced superconducting gap at $B=0$, and $\mu$ is the chemical potential in the quantum wire. In the limits of zero temperature, hard induced gap and weak tunnel coupling, MBS manifest as a $2e^2/h$ peak in conductance at zero tunneling bias. The peak emerges after the bulk superconducting gap in the nanowire closes and re-opens. Beyond the re-opening point the peak is robust - meaning it does not deviate from zero bias until superconductivity is fully suppressed by external field or another subband crosses the Fermi level. MBS come in pairs, and therefore a correlated zero-bias peak should be observed on the opposite end of the nanowire. Spin-orbit anisotropy implies that zero-bias peaks should vanish for a specific magnetic field orientation that is collinear with the effective spin-orbit field. 

The demonstration of all of the above basic tunneling predictions in the same nanowire will likely amount to proof of  Majorana bound states beyond reasonable doubt. To date, this has not been possible, despite steady progress in growth and fabrication \cite{krogstrup2015, Shabani2016,Gazibegovicnature2017}. A given device can be tuned to display one or more tunneling signatures of MBS but not to simultaneously confirm all of the basic expectations. The discrepancies may still be consistent with MBS but ascribed to experimental limitations such as finite temperature, soft induced gap, disorder, short nanowire length, critical field anisotropy. When experimental limiatations are accounted for, MBS are expected to result in conductance peak oscillations around zero bias, reduced peak height, no gap closing and/or re-opening and distorted topological phase boundary. As noted above, several theories and experiments furthermore point out that these features are shared between imperfect MBS and trivial ABS making the two effects challenging to distinguish. In this manuscript we study the phenomenology of low-bias resonances without assuming MBS, but with a goal of deeper understanding the superconductor-semiconductor hybrid system.

\begin{figure}[b!]
\centering
  \includegraphics[width=0.45\textwidth]{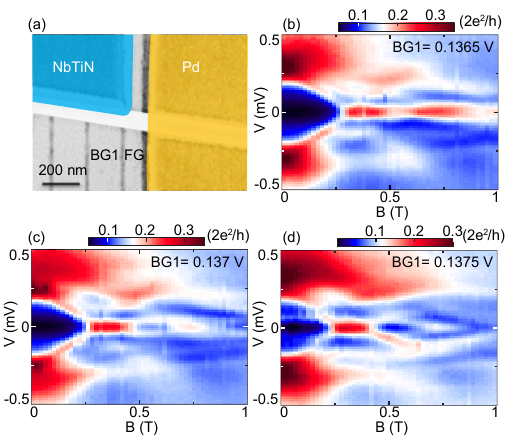}
  \caption { (a) Scanning electron micrograph of the studied device\textcolor{red}{(Device A)}. The bottom gates FG (100 nm wide) and BG1 (200 nm wide) are made of Ti(5~nm)/Au(10~nm). The nanowire is about 100 nm in diameter. The superconducting contact is a trilayer of Ti(5~nm)/NbTi(5~nm)/NbTiN(150~nm), while the normal contact is a Ti(15~nm)/Pd(150~nm) stack. (b-d) Differential conductance maps in bias voltage V versus magnetic field at $BG1= 0.1365$, $0.137$ and $0.1375$~V, respectively. \textcolor{red}{$FG$ is fixed at $0.53~V$ .} }
 \label{fig1}
\end{figure}

\begin{figure*}[t!]
\centering
  \includegraphics[width=0.7\textwidth]{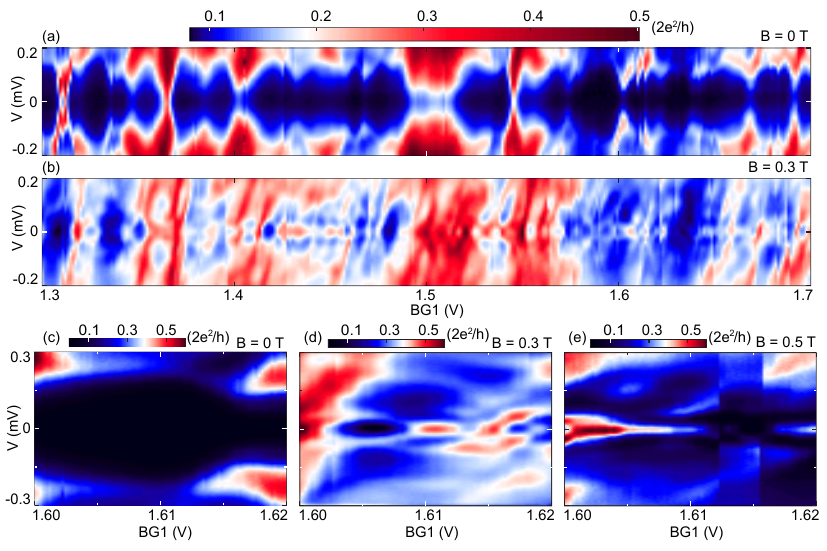}
\caption{
Ubiquitous ZBP in extended range of gate $BG1$ \textcolor{red}{(Device B)}. (a-b) Differentiate conductance maps in bias voltage V versus $BG1$ at $B= 0$ and $0.3$~T, respectively. (c-e) Differentiate conductance maps in bias voltage $V$ versus $BG1$ in a small range at $B= 0$, $0.3$ and $0.5$~T, respectively. \textcolor{red}{$FG$ is fixed at $0.45~V$ .}
 \label{fig2}}
\end{figure*}

Devices are fabricated using InSb semiconductor nanowires with NbTiN contacts (nominally identical to that used in \cite{JChen2017}, though device-to-device variations are common (Fig.\ref{fig1}(a), see supplemental materials for additional devices). Prior to the deposition of NbTiN, sulfur passivation is carried out followed by a gentle Ar plasma cleaning in order to obtain a transparent superconductor/semiconductor interface. A normal metal Pd contact is then fabricated to perform tunneling spectroscopy by varying bias voltage $V$ between normal and superconducting contacts. Electrical measurements are performed in a dilution refrigerator at a base temperature of $30$ mK, by a standard low-frequency lock-in technique (See detailed measurement conditions in Supplemental Material).

The electrostatic coupling of gates to the nanowire is enhanced due to half-coverage of the nanowire by the superconductor, as well due to the use of a thin layer of high-$\kappa$ gate dielectric (HfO$_2$, 10~nm). The gate effect is much stronger than in fully-covered nanowires \cite{gulnatnano2018}, or where side gates and/or thicker dielectric layers are used \cite{MourikScience2012, AlbrechtNature2016}. Stronger electrostatic coupling allows us to tune the density underneath the superconductor in a wider range, and observe a larger variety of subgap states as shown below. On the flip side, partial coverage may result in weaker induced superconductivity and soft gap \cite{VuikNJP2016}. Following a standard procedure for Majorana experiments \cite{MourikScience2012}, we create a single tunnel barrier between the normal and superconducting contacts by tuning gate $FG$ (once set $FG$ remains fixed). The gates left of $BG1$ are set to large negative voltages ($-2.5~$V) and not changed during the measurements. Those gates have no significant effect on the subgap states studied here.

We explore the magnetic field evolution of tunneling conductance in Figs. \ref{fig1}(b)-(d). At zero field, this device exhibits a soft but otherwise featureless superconducting gap characterized by smooth evolution of suppressed conductance within the gap as a function of bias. Such soft gap presents a decoherence pathway for futuristic topological qubits but it does not prevent us from studying the subgap spectroscopy here. In Fig.\ref{fig1}(b), the evolution within the magnetic field range $0-300$~mT looks like a closing of the induced gap: the suppressed conductance window around zero bias shrinks and two branches of high conductance move from the apparent induced gap edges ($V = \pm 250 \mu V$) toward lower bias reaching zero bias at around 300~mT. Beyond $B=300$~mT, an apparent zero bias resonance is observed over a significant range of magnetic field, up to at least $B = 1$~T. This range, expressed in Zeeman energy using a lower bound on InSb g-factor of 30, greatly exceeds the bias width of that resonance---thus we identify it as `pinned' to zero energy (line traces in Fig. S1 of the Supplemental Materials).

Fig. \ref{fig1}(c) shows that with a minor variation in $BG1$ a single zero-bias resonance can be transformed into a pair of low-bias resonances oscillating around zero bias as magnetic field is increased to $1$~T (up to $2$~T in Supplemental Material). Such oscillations are consistent with MBS in a short nanowire \cite{StanescuPRB2013}, and in fact data in Fig. \ref{fig1} (b) can also be interpreted as similar oscillations of smaller amplitude, less than the resonance width. Fig. \ref{fig1}(d), however, conveys a different picture. After another change in $BG1$ that should not alter the bulk density in any significant way, we can resolve that the apparent oscillations are actually superimposed of two unrelated pairs of resonances moving to zero bias at different magnetic fields, $0.4$~T and $0.7$~T. This demonstrates that the visibility of different branches can be strongly affected by minor changes in gate voltages, and some of the branches may become invisible in color maps, creating the appearance of a sole zero bias resonance or a pair of oscillating resonances, both being important signatures of MBS.

The ubiquity of zero-bias features like those in Fig.\ref{fig1} is demonstrated in Fig.\ref{fig2} \textcolor{red}{(Device B)}. Because of the extended range of $BG1$ shown and because of the strong electrostatic coupling of $BG1$ to the nanowire, a large number of transient resonances can be seen crisscrossing the subgap region without sticking to zero bias at zero field (Fig.\ref{fig2}(a)). These are due to states localized near the tunneling barrier. At finite magnetic field $B = 0.3$~T, the transient resonances are still visible, but another set of features tightly confined close to zero bias is now observed throughout the presented range of $BG1$ (Fig.\ref{fig2}(b)). Close to 30 distinct ZBP regions are observed. If all of these ZBPs were due to topological superconductivity, we would expect being able to tune through tens of 1D subbands, which is inconsistent with quantum point contact measurements on similar nanowires \cite{vanweperennanolett13}. Data in Fig.\ref{fig2}(b) are similar to barrier gate scans in Mourik et al. \cite{MourikScience2012}, which used the same nanowires and superconductors, though a different gate layout with a weaker $BG1$ coupling. 

We zoom in on a representative $BG1$ range in Figs.\ref{fig2}(c)-(e). At zero field the inside of the induced gap for $|V| < 250 \mu V$ is featureless on this scale (Fig.\ref{fig2}(c)). In the same gate range at finite field $B = 0.3$~T (Fig.\ref{fig2}(d)), three oscillations around zero bias and higher bias subgap states are observed. At a higher field $B = 0.5$~T (Fig. \ref{fig2}(e)), an extended zero-bias peak is observed. Over a range of $BG$ between $1.61$~V and $1.62$~V the ZBP vanishes, however this is an artifact due to  charge jumps, i.e. charge rearrangements near the gate leading to a momentary shift in the electrostatic potential. Such charge jumps are also ubiquitous and appear in many published results \cite{ZhangNature2018}. 

\begin{figure}[t!]
\centering
  \includegraphics[width=0.45\textwidth]{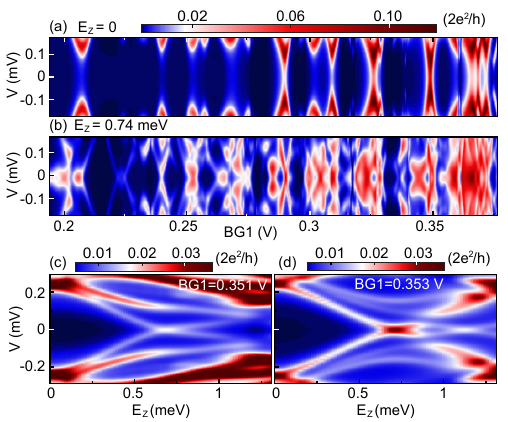}
\caption{
Calculated differential conductance as a function of $BG1$ and bias-voltage $V$ for (a) $E_Z=0$ and (b) $E_Z=0.74~$meV.
The FG voltage is $0.38~$V and the temperature $k_B T=0.15~$meV. (c-d) Calculated differential conductance as a function of Zeeman energy and bias voltage $V$ for $BG1= 0.351$ and $0.353~$V, respectively.
 \label{fig3}}
\end{figure}

We observe that the near-zero bias states often merge continuously into the transient resonances above the induced gap. This implies a relation between the two types of features. This behavior is expected in quantum dots strongly coupled to superconductors, where transport resonances due to ABS split from and merge into the induced gap as the dot occupation changes from even to odd \cite{eichlerprl2007, LeeNatnano2014}. In this framework, the regime in Fig.\ref{fig2} is consistent with several coupled quantum dots formed near the superconductor. Note that the absence of Coulomb blockade suggests open quantum dots and transparent contact to the superconductor. The open dots may be connected both in series and in parallel.

To model our devices we perform 3D Schr\"{o}dinger-Poisson calculations that incorporate geometric and electrostatic details of the experimental device \cite{Woods2018a}. The calculations naturally capture the multi-band nature of the system and its highly inhomogeneous electrostatic potential, which turn out to be the crucial elements responsible for the ubiquitous zero bias peaks. The inhomogeneity arises due to device geometry, while disorder is not included in the model. A detailed description of the model can be found in the conjoint theoretical paper \cite{Woods2019}.

First, we demonstrate that the model generates ubiquitous zero bias peaks, as seen in the experiment, by calculating the differential conductance \cite{BlonderPhysRevB1982} as a function of the $BG1$ voltage. The results are shown in Fig. \ref{fig3} (compare with  Fig.\ref{fig2}). At zero magnetic field (Fig. \ref{fig3}(a)), the differential conductance is characterized by multiple sub-gap resonances that approach or cross zero bias without sticking. At finite field (Fig. \ref{fig3}(b)), one notices features that are confined near zero energy. Examples of differential conductance maps as a function of Zeeman energy and bias are shown in Figs. \ref{fig3}(c)-(d). In Fig. \ref{fig3}(c), we notice an in-gap mode that collapses to zero energy at $E_Z\approx 0.7~$meV, then splits at higher $E_Z$. A slight change in $BG1$ generates a low-energy mode that remains near zero bias over a large range of $E_z$ (Fig. \ref{fig3}(d)). 

Next, we address the key question regarding the nature of the low-energy states by studying the band and real-space structure of the corresponding wave functions (also see the Supplemental Material). We find that the ubiquitous low-energy states are not MBS emerging in a segment of the wire, or partially-separated MBS induced by soft confinement, but rather ABS pinned near zero energy by level repulsion. As detailed in the conjoint theory paper \cite{Woods2019}, inter-band coupling can give rise to ABS that stick near zero energy due to anti-crossings between multiple modes approaching zero energy at different magnetic fields. For example, in Fig. \ref{fig3}(d) one can  distinguish two low-energy modes that cross zero energy at $E_Z\approx 0.7~$meV and $E_Z\approx 1.1~$meV, respectively, displaying an anti-crossing behavior (near $E_Z\approx 0.9~$meV). Evidence of similar level repulsion behavior can be found in the experimental results shown in Fig. {\ref{fig1}}(b-c). We note that the inter-band coupling arises from the evolution along the length of the wire of the transverse profiles of the various bands due to the electrostatic potential nonuniformity.
As explained in the theory paper \cite{Woods2019}, single-subband models cannot capture this zero-bias pinning behavior for short nanowire segments of 200 nm (the width of $BG1$) without assuming overlapping MBS.

\begin{figure}[t!]
\centering
  \includegraphics[width=0.45\textwidth]{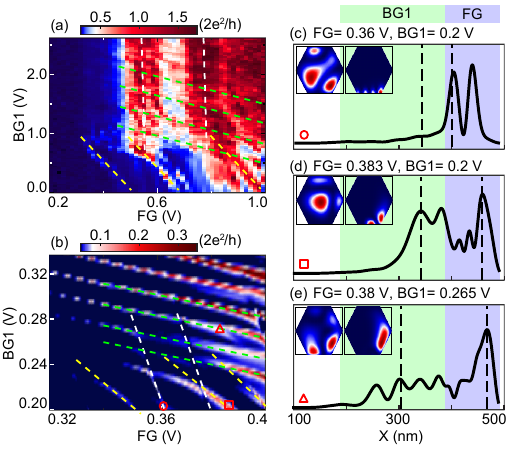}
\caption{
(a) Measured and (b) calculated zero-bias voltage conductance at zero magnetic field as a function of $FG$ and $BG1$. Note the three types of resonances characterized by different slopes (dashed lines are guides for the eye). (c-e) Calculated wavefunction profiles for the states marked in panel (b).  The insets show the transverse profiles at the locations marked by black dashed lines. Regions of $BG1$ and $FG$ are marked by green and purple shadows, respectively.
 \label{fig4}}
\end{figure}

We investigate the spatial characteristics of the low-energy states at zero magnetic field by mapping the zero bias conductance as a function of $BG1$ and $FG$. The experimental results are shown in  Fig. \ref{fig4}(a), while the numerical results are given in  Fig. \ref{fig4}(b). The remarkable common feature is the presence of three types of resonances characterized by different slopes, which we attribute to distinct families of low-energy states having different couplings to the gate potentials. The nearly-vertical resonances in Fig.\ref{fig4}(a) are generated by states coupled primarily to the $FG$ gate (white dashed lines). These resonances are also seen in the calculation (Fig.\ref{fig4}(b)), although their amplitude is reduced (see also the Supplemental Materials). The wavefunction profile of a typical state associated with this type of resonance (Fig.\ref{fig4}(c)) reveals that most of its weight is located in the $FG$ region (see inset). The two additional sets of resonances are generated by states electrostatically coupled to both $FG$ and $BG1$ (green and yellow). As revealed by the wave function profiles shown in Figs.\ref{fig4}(d)-(e), these states have significant weight in both $BG1$ and $FG$ regions. However, the transverse profiles (see insets) show that the state in Fig.\ref{fig4}(e) is located closer to $BG1$ and farther away from $FG$ as compared to the state in Fig.\ref{fig4}(d), which explains the different slopes of the corresponding resonances. The existence of these distinct families of states demonstrates that the low-energy physics is controlled by modes localized in different adjacent regions (i.e. the $FG$ region and the covered and uncovered $BG1$ regions) that are coupled to one another. At finite magnetic field, this generically produces low-energy ABS resonances pinned near zero energy through the inter-band coupling mechanism discussed in detail in Ref. \cite{Woods2019}. We note a discrepancy in gate voltage and conductance scales between Fig.\ref{fig4}(a) and Fig.\ref{fig4}(b), likely as a result of device-dependent gate screening variations and high sensitivity of conductance to tunneling rates.

In conclusion, we have demonstrated that many of the commonly discussed features of MBS in nanowires, such as gap closing, zero-bias pinning in magnetic field or gate, and peak oscillations around zero bias, are ubiquitous and easily observed when ensembles of trivial ABS are present. Evidence of MBS in tunneling experiments should therefore be accompanied by detailed studies of  subgap resonances in the extended gate voltage range. For example, an earlier study of a similar device has revealed zero-bias peaks occupying a large continuous region of field-gate space with a boundary similar to the basic topological condition \cite{JChen2017}.

Nevertheless, since tunneling measurements have so far not yielded a definite MBS proof, it is intuitively attractive to  explore more sophisticated techniques, e.g. the fractional Josephson effect \cite{RokhinsonNPhys2012}, Majorana fusion or even braiding \cite{AasenPRX2016}. However, the added measurement complexity will not help resolve the experimental limitations of the tunneling experiments, since the limitations remain rooted in the growth and fabrication. It is also unclear whether advanced techniques can reveal signatures unique to MBS, and whether they are better at distinguishing MBS from ABS \cite{houzetprl2013}. At the same time, tunneling remains powerful in surveying the subgap spectra in proximitized nanowires, thereby guiding device design and fabrication towards a more ideal regime in which MBS can be  demonstrated unambiguously.

We thank V. Mourik and K. Zuo for comments on the manuscript. T.D.S. acknowledges  NSF DMR-1414683. S.M.F. acknowledges NSF DMR-1743972, NSF PIRE-1743717, ONR and ARO.

\bibliographystyle{apsrev4-1}
\bibliography{CHEN_ZBP_Ref.bib}

\begin{thebibliography}{44}%
\makeatletter
\providecommand \@ifxundefined [1]{%
 \@ifx{#1\undefined}
}%
\providecommand \@ifnum [1]{%
 \ifnum #1\expandafter \@firstoftwo
 \else \expandafter \@secondoftwo
 \fi
}%
\providecommand \@ifx [1]{%
 \ifx #1\expandafter \@firstoftwo
 \else \expandafter \@secondoftwo
 \fi
}%
\providecommand \natexlab [1]{#1}%
\providecommand \enquote  [1]{``#1''}%
\providecommand \bibnamefont  [1]{#1}%
\providecommand \bibfnamefont [1]{#1}%
\providecommand \citenamefont [1]{#1}%
\providecommand \href@noop [0]{\@secondoftwo}%
\providecommand \href [0]{\begingroup \@sanitize@url \@href}%
\providecommand \@href[1]{\@@startlink{#1}\@@href}%
\providecommand \@@href[1]{\endgroup#1\@@endlink}%
\providecommand \@sanitize@url [0]{\catcode `\\12\catcode `\$12\catcode `\&12\catcode `\#12\catcode `\^12\catcode `\_12\catcode `\%12\relax}%
\providecommand \@@startlink[1]{}%
\providecommand \@@endlink[0]{}%
\providecommand \url  [0]{\begingroup\@sanitize@url \@url }%
\providecommand \@url [1]{\endgroup\@href {#1}{\urlprefix }}%
\providecommand \urlprefix  [0]{URL }%
\providecommand \Eprint [0]{\href }%
\providecommand \doibase [0]{http://dx.doi.org/}%
\providecommand \selectlanguage [0]{\@gobble}%
\providecommand \bibinfo  [0]{\@secondoftwo}%
\providecommand \bibfield  [0]{\@secondoftwo}%
\providecommand \translation [1]{[#1]}%
\providecommand \BibitemOpen [0]{}%
\providecommand \bibitemStop [0]{}%
\providecommand \bibitemNoStop [0]{.\EOS\space}%
\providecommand \EOS [0]{\spacefactor3000\relax}%
\providecommand \BibitemShut  [1]{\csname bibitem#1\endcsname}%
\let\auto@bib@innerbib\@empty
\bibitem [{\citenamefont {Fu}\ and\ \citenamefont {Kane}(2008)}]{Fu2008PRL}%
  \BibitemOpen
  \bibfield  {author} {\bibinfo {author} {\bibfnamefont {L.}~\bibnamefont {Fu}}\ and\ \bibinfo {author} {\bibfnamefont {C.~L.}\ \bibnamefont {Kane}},\ }\href {\doibase 10.1103/PhysRevLett.100.096407} {\bibfield  {journal} {\bibinfo  {journal} {Phys. Rev. Lett.}\ }\textbf {\bibinfo {volume} {100}},\ \bibinfo {pages} {096407} (\bibinfo {year} {2008})}\BibitemShut {NoStop}%
\bibitem [{\citenamefont {Alicea}(2010)}]{AliceaPRB2010}%
  \BibitemOpen
  \bibfield  {author} {\bibinfo {author} {\bibfnamefont {J.}~\bibnamefont {Alicea}},\ }\href {\doibase 10.1103/PhysRevB.81.125318} {\bibfield  {journal} {\bibinfo  {journal} {Phys. Rev. B}\ }\textbf {\bibinfo {volume} {81}},\ \bibinfo {pages} {125318} (\bibinfo {year} {2010})}\BibitemShut {NoStop}%
\bibitem [{\citenamefont {Sau}\ \emph {et~al.}(2010)\citenamefont {Sau}, \citenamefont {Tewari}, \citenamefont {Lutchyn}, \citenamefont {Stanescu},\ and\ \citenamefont {Das~Sarma}}]{SauPRB2010}%
  \BibitemOpen
  \bibfield  {author} {\bibinfo {author} {\bibfnamefont {J.~D.}\ \bibnamefont {Sau}}, \bibinfo {author} {\bibfnamefont {S.}~\bibnamefont {Tewari}}, \bibinfo {author} {\bibfnamefont {R.~M.}\ \bibnamefont {Lutchyn}}, \bibinfo {author} {\bibfnamefont {T.~D.}\ \bibnamefont {Stanescu}}, \ and\ \bibinfo {author} {\bibfnamefont {S.}~\bibnamefont {Das~Sarma}},\ }\href {\doibase 10.1103/PhysRevB.82.214509} {\bibfield  {journal} {\bibinfo  {journal} {Phys. Rev. B}\ }\textbf {\bibinfo {volume} {82}},\ \bibinfo {pages} {214509} (\bibinfo {year} {2010})}\BibitemShut {NoStop}%
\bibitem [{\citenamefont {Lutchyn}\ \emph {et~al.}(2010)\citenamefont {Lutchyn}, \citenamefont {Sau},\ and\ \citenamefont {Das~Sarma}}]{LutchynPRL2010}%
  \BibitemOpen
  \bibfield  {author} {\bibinfo {author} {\bibfnamefont {R.~M.}\ \bibnamefont {Lutchyn}}, \bibinfo {author} {\bibfnamefont {J.~D.}\ \bibnamefont {Sau}}, \ and\ \bibinfo {author} {\bibfnamefont {S.}~\bibnamefont {Das~Sarma}},\ }\href@noop {} {\bibfield  {journal} {\bibinfo  {journal} {Physical review letters}\ }\textbf {\bibinfo {volume} {105}},\ \bibinfo {pages} {077001} (\bibinfo {year} {2010})}\BibitemShut {NoStop}%
\bibitem [{\citenamefont {Oreg}\ \emph {et~al.}(2010)\citenamefont {Oreg}, \citenamefont {Refael},\ and\ \citenamefont {von Oppen}}]{OregPRL2010}%
  \BibitemOpen
  \bibfield  {author} {\bibinfo {author} {\bibfnamefont {Y.}~\bibnamefont {Oreg}}, \bibinfo {author} {\bibfnamefont {G.}~\bibnamefont {Refael}}, \ and\ \bibinfo {author} {\bibfnamefont {F.}~\bibnamefont {von Oppen}},\ }\href@noop {} {\bibfield  {journal} {\bibinfo  {journal} {Physical review letters}\ }\textbf {\bibinfo {volume} {105}},\ \bibinfo {pages} {177002} (\bibinfo {year} {2010})}\BibitemShut {NoStop}%
\bibitem [{\citenamefont {Alicea}(2012)}]{AliceaPRP2012}%
  \BibitemOpen
  \bibfield  {author} {\bibinfo {author} {\bibfnamefont {J.}~\bibnamefont {Alicea}},\ }\href {http://stacks.iop.org/0034-4885/75/i=7/a=076501} {\bibfield  {journal} {\bibinfo  {journal} {Reports on Progress in Physics}\ }\textbf {\bibinfo {volume} {75}},\ \bibinfo {pages} {076501} (\bibinfo {year} {2012})}\BibitemShut {NoStop}%
\bibitem [{\citenamefont {Beenakker}(2013)}]{BeenakkerARCMP2013}%
  \BibitemOpen
  \bibfield  {author} {\bibinfo {author} {\bibfnamefont {C.}~\bibnamefont {Beenakker}},\ }\href {\doibase 10.1146/annurev-conmatphys-030212-184337} {\bibfield  {journal} {\bibinfo  {journal} {Annual Review of Condensed Matter Physics}\ }\textbf {\bibinfo {volume} {4}},\ \bibinfo {pages} {113} (\bibinfo {year} {2013})},\ \Eprint {http://arxiv.org/abs/http://dx.doi.org/10.1146/annurev-conmatphys-030212-184337} {http://dx.doi.org/10.1146/annurev-conmatphys-030212-184337} \BibitemShut {NoStop}%
\bibitem [{\citenamefont {Read}\ and\ \citenamefont {Green}(2000)}]{ReadPRB2000}%
  \BibitemOpen
  \bibfield  {author} {\bibinfo {author} {\bibfnamefont {N.}~\bibnamefont {Read}}\ and\ \bibinfo {author} {\bibfnamefont {D.}~\bibnamefont {Green}},\ }\href@noop {} {\bibfield  {journal} {\bibinfo  {journal} {Physical Review B}\ }\textbf {\bibinfo {volume} {61}},\ \bibinfo {pages} {10267} (\bibinfo {year} {2000})}\BibitemShut {NoStop}%
\bibitem [{\citenamefont {Mourik}\ \emph {et~al.}(2012)\citenamefont {Mourik}, \citenamefont {Zuo}, \citenamefont {Frolov}, \citenamefont {Plissard}, \citenamefont {Bakkers},\ and\ \citenamefont {Kouwenhoven}}]{MourikScience2012}%
  \BibitemOpen
  \bibfield  {author} {\bibinfo {author} {\bibfnamefont {V.}~\bibnamefont {Mourik}}, \bibinfo {author} {\bibfnamefont {K.}~\bibnamefont {Zuo}}, \bibinfo {author} {\bibfnamefont {S.~M.}\ \bibnamefont {Frolov}}, \bibinfo {author} {\bibfnamefont {S.}~\bibnamefont {Plissard}}, \bibinfo {author} {\bibfnamefont {E.~P.}\ \bibnamefont {Bakkers}}, \ and\ \bibinfo {author} {\bibfnamefont {L.~P.}\ \bibnamefont {Kouwenhoven}},\ }\href@noop {} {\bibfield  {journal} {\bibinfo  {journal} {Science}\ }\textbf {\bibinfo {volume} {336}},\ \bibinfo {pages} {1003} (\bibinfo {year} {2012})}\BibitemShut {NoStop}%
\bibitem [{\citenamefont {Das}\ \emph {et~al.}(2012)\citenamefont {Das}, \citenamefont {Ronen}, \citenamefont {Most}, \citenamefont {Oreg}, \citenamefont {Heiblum},\ and\ \citenamefont {Shtrikman}}]{DasNatphys2012}%
  \BibitemOpen
  \bibfield  {author} {\bibinfo {author} {\bibfnamefont {A.}~\bibnamefont {Das}}, \bibinfo {author} {\bibfnamefont {Y.}~\bibnamefont {Ronen}}, \bibinfo {author} {\bibfnamefont {Y.}~\bibnamefont {Most}}, \bibinfo {author} {\bibfnamefont {Y.}~\bibnamefont {Oreg}}, \bibinfo {author} {\bibfnamefont {M.}~\bibnamefont {Heiblum}}, \ and\ \bibinfo {author} {\bibfnamefont {H.}~\bibnamefont {Shtrikman}},\ }\href@noop {} {\bibfield  {journal} {\bibinfo  {journal} {Nature Physics}\ }\textbf {\bibinfo {volume} {8}},\ \bibinfo {pages} {887} (\bibinfo {year} {2012})}\BibitemShut {NoStop}%
\bibitem [{\citenamefont {Deng}\ \emph {et~al.}(2012)\citenamefont {Deng}, \citenamefont {Yu}, \citenamefont {Huang}, \citenamefont {Larsson}, \citenamefont {Caroff},\ and\ \citenamefont {Xu}}]{DengNanolett2012}%
  \BibitemOpen
  \bibfield  {author} {\bibinfo {author} {\bibfnamefont {M.}~\bibnamefont {Deng}}, \bibinfo {author} {\bibfnamefont {C.}~\bibnamefont {Yu}}, \bibinfo {author} {\bibfnamefont {G.}~\bibnamefont {Huang}}, \bibinfo {author} {\bibfnamefont {M.}~\bibnamefont {Larsson}}, \bibinfo {author} {\bibfnamefont {P.}~\bibnamefont {Caroff}}, \ and\ \bibinfo {author} {\bibfnamefont {H.}~\bibnamefont {Xu}},\ }\href@noop {} {\bibfield  {journal} {\bibinfo  {journal} {Nano letters}\ }\textbf {\bibinfo {volume} {12}},\ \bibinfo {pages} {6414} (\bibinfo {year} {2012})}\BibitemShut {NoStop}%
\bibitem [{\citenamefont {Finck}\ \emph {et~al.}(2013)\citenamefont {Finck}, \citenamefont {Van~Harlingen}, \citenamefont {Mohseni}, \citenamefont {Jung},\ and\ \citenamefont {Li}}]{FinckPRL2013}%
  \BibitemOpen
  \bibfield  {author} {\bibinfo {author} {\bibfnamefont {A.~D.~K.}\ \bibnamefont {Finck}}, \bibinfo {author} {\bibfnamefont {D.~J.}\ \bibnamefont {Van~Harlingen}}, \bibinfo {author} {\bibfnamefont {P.~K.}\ \bibnamefont {Mohseni}}, \bibinfo {author} {\bibfnamefont {K.}~\bibnamefont {Jung}}, \ and\ \bibinfo {author} {\bibfnamefont {X.}~\bibnamefont {Li}},\ }\href {\doibase 10.1103/PhysRevLett.110.126406} {\bibfield  {journal} {\bibinfo  {journal} {Phys. Rev. Lett.}\ }\textbf {\bibinfo {volume} {110}},\ \bibinfo {pages} {126406} (\bibinfo {year} {2013})}\BibitemShut {NoStop}%
\bibitem [{\citenamefont {Churchill}\ \emph {et~al.}(2013)\citenamefont {Churchill}, \citenamefont {Fatemi}, \citenamefont {Grove-Rasmussen}, \citenamefont {Deng}, \citenamefont {Caroff}, \citenamefont {Xu},\ and\ \citenamefont {Marcus}}]{ChurchillPRB2013}%
  \BibitemOpen
  \bibfield  {author} {\bibinfo {author} {\bibfnamefont {H.~O.~H.}\ \bibnamefont {Churchill}}, \bibinfo {author} {\bibfnamefont {V.}~\bibnamefont {Fatemi}}, \bibinfo {author} {\bibfnamefont {K.}~\bibnamefont {Grove-Rasmussen}}, \bibinfo {author} {\bibfnamefont {M.}~\bibnamefont {Deng}}, \bibinfo {author} {\bibfnamefont {P.}~\bibnamefont {Caroff}}, \bibinfo {author} {\bibfnamefont {H.}~\bibnamefont {Xu}}, \ and\ \bibinfo {author} {\bibfnamefont {C.~M.}\ \bibnamefont {Marcus}},\ }\href@noop {} {\bibfield  {journal} {\bibinfo  {journal} {Physical Review B}\ }\textbf {\bibinfo {volume} {87}},\ \bibinfo {pages} {241401} (\bibinfo {year} {2013})}\BibitemShut {NoStop}%
\bibitem [{\citenamefont {Nadj-Perge}\ \emph {et~al.}(2014)\citenamefont {Nadj-Perge}, \citenamefont {Drozdov}, \citenamefont {Li}, \citenamefont {Chen}, \citenamefont {Jeon}, \citenamefont {Seo}, \citenamefont {MacDonald}, \citenamefont {Bernevig},\ and\ \citenamefont {Yazdani}}]{Nadj-PergeScience2014}%
  \BibitemOpen
  \bibfield  {author} {\bibinfo {author} {\bibfnamefont {S.}~\bibnamefont {Nadj-Perge}}, \bibinfo {author} {\bibfnamefont {I.~K.}\ \bibnamefont {Drozdov}}, \bibinfo {author} {\bibfnamefont {J.}~\bibnamefont {Li}}, \bibinfo {author} {\bibfnamefont {H.}~\bibnamefont {Chen}}, \bibinfo {author} {\bibfnamefont {S.}~\bibnamefont {Jeon}}, \bibinfo {author} {\bibfnamefont {J.}~\bibnamefont {Seo}}, \bibinfo {author} {\bibfnamefont {A.~H.}\ \bibnamefont {MacDonald}}, \bibinfo {author} {\bibfnamefont {B.~A.}\ \bibnamefont {Bernevig}}, \ and\ \bibinfo {author} {\bibfnamefont {A.}~\bibnamefont {Yazdani}},\ }\href@noop {} {\bibfield  {journal} {\bibinfo  {journal} {Science}\ }\textbf {\bibinfo {volume} {346}},\ \bibinfo {pages} {602} (\bibinfo {year} {2014})}\BibitemShut {NoStop}%
\bibitem [{\citenamefont {Albrecht}\ \emph {et~al.}(2016{\natexlab{a}})\citenamefont {Albrecht}, \citenamefont {Higginbotham}, \citenamefont {Madsen}, \citenamefont {Kuemmeth}, \citenamefont {Jespersen}, \citenamefont {Nyg{\aa}rd}, \citenamefont {Krogstrup},\ and\ \citenamefont {Marcus}}]{Albrecht2016Nature}%
  \BibitemOpen
  \bibfield  {author} {\bibinfo {author} {\bibfnamefont {S.~M.}\ \bibnamefont {Albrecht}}, \bibinfo {author} {\bibfnamefont {A.~P.}\ \bibnamefont {Higginbotham}}, \bibinfo {author} {\bibfnamefont {M.}~\bibnamefont {Madsen}}, \bibinfo {author} {\bibfnamefont {F.}~\bibnamefont {Kuemmeth}}, \bibinfo {author} {\bibfnamefont {T.~S.}\ \bibnamefont {Jespersen}}, \bibinfo {author} {\bibfnamefont {J.}~\bibnamefont {Nyg{\aa}rd}}, \bibinfo {author} {\bibfnamefont {P.}~\bibnamefont {Krogstrup}}, \ and\ \bibinfo {author} {\bibfnamefont {C.~M.}\ \bibnamefont {Marcus}},\ }\href {https://doi.org/10.1038/nature17162} {\bibfield  {journal} {\bibinfo  {journal} {Nature}\ }\textbf {\bibinfo {volume} {531}},\ \bibinfo {pages} {206 EP } (\bibinfo {year} {2016}{\natexlab{a}})}\BibitemShut {NoStop}%
\bibitem [{\citenamefont {Deng}\ \emph {et~al.}(2016)\citenamefont {Deng}, \citenamefont {Vaitiekenas}, \citenamefont {Hansen}, \citenamefont {Danon}, \citenamefont {Leijnse}, \citenamefont {Flensberg}, \citenamefont {Nyg{\r a}rd}, \citenamefont {Krogstrup},\ and\ \citenamefont {Marcus}}]{DengScience2016}%
  \BibitemOpen
  \bibfield  {author} {\bibinfo {author} {\bibfnamefont {M.~T.}\ \bibnamefont {Deng}}, \bibinfo {author} {\bibfnamefont {S.}~\bibnamefont {Vaitiekenas}}, \bibinfo {author} {\bibfnamefont {E.~B.}\ \bibnamefont {Hansen}}, \bibinfo {author} {\bibfnamefont {J.}~\bibnamefont {Danon}}, \bibinfo {author} {\bibfnamefont {M.}~\bibnamefont {Leijnse}}, \bibinfo {author} {\bibfnamefont {K.}~\bibnamefont {Flensberg}}, \bibinfo {author} {\bibfnamefont {J.}~\bibnamefont {Nyg{\r a}rd}}, \bibinfo {author} {\bibfnamefont {P.}~\bibnamefont {Krogstrup}}, \ and\ \bibinfo {author} {\bibfnamefont {C.~M.}\ \bibnamefont {Marcus}},\ }\href {\doibase 10.1126/science.aaf3961} {\bibfield  {journal} {\bibinfo  {journal} {Science}\ }\textbf {\bibinfo {volume} {354}},\ \bibinfo {pages} {1557} (\bibinfo {year} {2016})}\BibitemShut {NoStop}%
\bibitem [{\citenamefont {Chen}\ \emph {et~al.}(2017)\citenamefont {Chen}, \citenamefont {Yu}, \citenamefont {Stenger}, \citenamefont {Hocevar}, \citenamefont {Car}, \citenamefont {Plissard}, \citenamefont {Bakkers}, \citenamefont {Stanescu},\ and\ \citenamefont {Frolov}}]{JChen2017}%
  \BibitemOpen
  \bibfield  {author} {\bibinfo {author} {\bibfnamefont {J.}~\bibnamefont {Chen}}, \bibinfo {author} {\bibfnamefont {P.}~\bibnamefont {Yu}}, \bibinfo {author} {\bibfnamefont {J.}~\bibnamefont {Stenger}}, \bibinfo {author} {\bibfnamefont {M.}~\bibnamefont {Hocevar}}, \bibinfo {author} {\bibfnamefont {D.}~\bibnamefont {Car}}, \bibinfo {author} {\bibfnamefont {S.~R.}\ \bibnamefont {Plissard}}, \bibinfo {author} {\bibfnamefont {E.~P. A.~M.}\ \bibnamefont {Bakkers}}, \bibinfo {author} {\bibfnamefont {T.~D.}\ \bibnamefont {Stanescu}}, \ and\ \bibinfo {author} {\bibfnamefont {S.~M.}\ \bibnamefont {Frolov}},\ }\href {\doibase 10.1126/sciadv.1701476} {\bibfield  {journal} {\bibinfo  {journal} {Science Advances}\ }\textbf {\bibinfo {volume} {3}} (\bibinfo {year} {2017}),\ 10.1126/sciadv.1701476}\BibitemShut {NoStop}%
\bibitem [{\citenamefont {G{\"u}l}\ \emph {et~al.}(2018)\citenamefont {G{\"u}l}, \citenamefont {Zhang}, \citenamefont {Bommer}, \citenamefont {de~Moor}, \citenamefont {Car}, \citenamefont {Plissard}, \citenamefont {Bakkers}, \citenamefont {Geresdi}, \citenamefont {Watanabe}, \citenamefont {Taniguchi} \emph {et~al.}}]{gulnatnano2018}%
  \BibitemOpen
  \bibfield  {author} {\bibinfo {author} {\bibfnamefont {{\"O}.}~\bibnamefont {G{\"u}l}}, \bibinfo {author} {\bibfnamefont {H.}~\bibnamefont {Zhang}}, \bibinfo {author} {\bibfnamefont {J.~D.}\ \bibnamefont {Bommer}}, \bibinfo {author} {\bibfnamefont {M.~W.}\ \bibnamefont {de~Moor}}, \bibinfo {author} {\bibfnamefont {D.}~\bibnamefont {Car}}, \bibinfo {author} {\bibfnamefont {S.~R.}\ \bibnamefont {Plissard}}, \bibinfo {author} {\bibfnamefont {E.~P.}\ \bibnamefont {Bakkers}}, \bibinfo {author} {\bibfnamefont {A.}~\bibnamefont {Geresdi}}, \bibinfo {author} {\bibfnamefont {K.}~\bibnamefont {Watanabe}}, \bibinfo {author} {\bibfnamefont {T.}~\bibnamefont {Taniguchi}},  \emph {et~al.},\ }\href@noop {} {\bibfield  {journal} {\bibinfo  {journal} {Nature nanotechnology}\ }\textbf {\bibinfo {volume} {13}},\ \bibinfo {pages} {192} (\bibinfo {year} {2018})}\BibitemShut {NoStop}%
\bibitem [{\citenamefont {Zhang}\ \emph {et~al.}(2018)\citenamefont {Zhang}, \citenamefont {Liu}, \citenamefont {Gazibegovic}, \citenamefont {Xu}, \citenamefont {Logan}, \citenamefont {Wang}, \citenamefont {van Loo}, \citenamefont {Bommer}, \citenamefont {de~Moor}, \citenamefont {Car}, \citenamefont {het Veld}, \citenamefont {van Veldhoven}, \citenamefont {Koelling}, \citenamefont {Verheijen}, \citenamefont {Pendharkar}, \citenamefont {Pennachio}, \citenamefont {Shojaei}, \citenamefont {Lee}, \citenamefont {Palmstrom}, \citenamefont {Bakkers}, \citenamefont {Sarma},\ and\ \citenamefont {Kouwenhoven}}]{ZhangNature2018}%
  \BibitemOpen
  \bibfield  {author} {\bibinfo {author} {\bibfnamefont {H.}~\bibnamefont {Zhang}}, \bibinfo {author} {\bibfnamefont {C.-X.}\ \bibnamefont {Liu}}, \bibinfo {author} {\bibfnamefont {S.}~\bibnamefont {Gazibegovic}}, \bibinfo {author} {\bibfnamefont {D.}~\bibnamefont {Xu}}, \bibinfo {author} {\bibfnamefont {J.~A.}\ \bibnamefont {Logan}}, \bibinfo {author} {\bibfnamefont {G.}~\bibnamefont {Wang}}, \bibinfo {author} {\bibfnamefont {N.}~\bibnamefont {van Loo}}, \bibinfo {author} {\bibfnamefont {J.~D.}\ \bibnamefont {Bommer}}, \bibinfo {author} {\bibfnamefont {M.~W.}\ \bibnamefont {de~Moor}}, \bibinfo {author} {\bibfnamefont {D.}~\bibnamefont {Car}}, \bibinfo {author} {\bibfnamefont {R.~L. M.~O.}\ \bibnamefont {het Veld}}, \bibinfo {author} {\bibfnamefont {P.~J.}\ \bibnamefont {van Veldhoven}}, \bibinfo {author} {\bibfnamefont {S.}~\bibnamefont {Koelling}}, \bibinfo {author} {\bibfnamefont {M.~A.}\ \bibnamefont {Verheijen}}, \bibinfo {author} {\bibfnamefont {M.}~\bibnamefont {Pendharkar}}, \bibinfo {author}
  {\bibfnamefont {D.~J.}\ \bibnamefont {Pennachio}}, \bibinfo {author} {\bibfnamefont {B.}~\bibnamefont {Shojaei}}, \bibinfo {author} {\bibfnamefont {J.~S.}\ \bibnamefont {Lee}}, \bibinfo {author} {\bibfnamefont {C.~J.}\ \bibnamefont {Palmstrom}}, \bibinfo {author} {\bibfnamefont {E.~P.}\ \bibnamefont {Bakkers}}, \bibinfo {author} {\bibfnamefont {S.~D.}\ \bibnamefont {Sarma}}, \ and\ \bibinfo {author} {\bibfnamefont {L.~P.}\ \bibnamefont {Kouwenhoven}},\ }\href {\doibase 10.1038/nature26142} {\bibfield  {journal} {\bibinfo  {journal} {Nature}\ }\textbf {\bibinfo {volume} {556}},\ \bibinfo {pages} {74} (\bibinfo {year} {2018})}\BibitemShut {NoStop}%
\bibitem [{\citenamefont {Suominen}\ \emph {et~al.}(2017)\citenamefont {Suominen}, \citenamefont {Kjaergaard}, \citenamefont {Hamilton}, \citenamefont {Shabani}, \citenamefont {Palmstr\o{}m}, \citenamefont {Marcus},\ and\ \citenamefont {Nichele}}]{SuominenPRL2017}%
  \BibitemOpen
  \bibfield  {author} {\bibinfo {author} {\bibfnamefont {H.~J.}\ \bibnamefont {Suominen}}, \bibinfo {author} {\bibfnamefont {M.}~\bibnamefont {Kjaergaard}}, \bibinfo {author} {\bibfnamefont {A.~R.}\ \bibnamefont {Hamilton}}, \bibinfo {author} {\bibfnamefont {J.}~\bibnamefont {Shabani}}, \bibinfo {author} {\bibfnamefont {C.~J.}\ \bibnamefont {Palmstr\o{}m}}, \bibinfo {author} {\bibfnamefont {C.~M.}\ \bibnamefont {Marcus}}, \ and\ \bibinfo {author} {\bibfnamefont {F.}~\bibnamefont {Nichele}},\ }\href {\doibase 10.1103/PhysRevLett.119.176805} {\bibfield  {journal} {\bibinfo  {journal} {Phys. Rev. Lett.}\ }\textbf {\bibinfo {volume} {119}},\ \bibinfo {pages} {176805} (\bibinfo {year} {2017})}\BibitemShut {NoStop}%
\bibitem [{\citenamefont {Nichele}\ \emph {et~al.}(2017)\citenamefont {Nichele}, \citenamefont {Drachmann}, \citenamefont {Whiticar}, \citenamefont {O'Farrell}, \citenamefont {Suominen}, \citenamefont {Fornieri}, \citenamefont {Wang}, \citenamefont {Gardner}, \citenamefont {Thomas}, \citenamefont {Hatke}, \citenamefont {Krogstrup}, \citenamefont {Manfra}, \citenamefont {Flensberg},\ and\ \citenamefont {Marcus}}]{NichelePRL2017}%
  \BibitemOpen
  \bibfield  {author} {\bibinfo {author} {\bibfnamefont {F.}~\bibnamefont {Nichele}}, \bibinfo {author} {\bibfnamefont {A.~C.~C.}\ \bibnamefont {Drachmann}}, \bibinfo {author} {\bibfnamefont {A.~M.}\ \bibnamefont {Whiticar}}, \bibinfo {author} {\bibfnamefont {E.~C.~T.}\ \bibnamefont {O'Farrell}}, \bibinfo {author} {\bibfnamefont {H.~J.}\ \bibnamefont {Suominen}}, \bibinfo {author} {\bibfnamefont {A.}~\bibnamefont {Fornieri}}, \bibinfo {author} {\bibfnamefont {T.}~\bibnamefont {Wang}}, \bibinfo {author} {\bibfnamefont {G.~C.}\ \bibnamefont {Gardner}}, \bibinfo {author} {\bibfnamefont {C.}~\bibnamefont {Thomas}}, \bibinfo {author} {\bibfnamefont {A.~T.}\ \bibnamefont {Hatke}}, \bibinfo {author} {\bibfnamefont {P.}~\bibnamefont {Krogstrup}}, \bibinfo {author} {\bibfnamefont {M.~J.}\ \bibnamefont {Manfra}}, \bibinfo {author} {\bibfnamefont {K.}~\bibnamefont {Flensberg}}, \ and\ \bibinfo {author} {\bibfnamefont {C.~M.}\ \bibnamefont {Marcus}},\ }\href {\doibase 10.1103/PhysRevLett.119.136803} {\bibfield  {journal}
  {\bibinfo  {journal} {Phys. Rev. Lett.}\ }\textbf {\bibinfo {volume} {119}},\ \bibinfo {pages} {136803} (\bibinfo {year} {2017})}\BibitemShut {NoStop}%
\bibitem [{\citenamefont {Lee}\ \emph {et~al.}(2012)\citenamefont {Lee}, \citenamefont {Jiang}, \citenamefont {Aguado}, \citenamefont {Katsaros}, \citenamefont {Lieber},\ and\ \citenamefont {De~Franceschi}}]{LeePRL2012}%
  \BibitemOpen
  \bibfield  {author} {\bibinfo {author} {\bibfnamefont {E.~J.~H.}\ \bibnamefont {Lee}}, \bibinfo {author} {\bibfnamefont {X.}~\bibnamefont {Jiang}}, \bibinfo {author} {\bibfnamefont {R.}~\bibnamefont {Aguado}}, \bibinfo {author} {\bibfnamefont {G.}~\bibnamefont {Katsaros}}, \bibinfo {author} {\bibfnamefont {C.~M.}\ \bibnamefont {Lieber}}, \ and\ \bibinfo {author} {\bibfnamefont {S.}~\bibnamefont {De~Franceschi}},\ }\href {\doibase 10.1103/PhysRevLett.109.186802} {\bibfield  {journal} {\bibinfo  {journal} {Phys. Rev. Lett.}\ }\textbf {\bibinfo {volume} {109}},\ \bibinfo {pages} {186802} (\bibinfo {year} {2012})}\BibitemShut {NoStop}%
\bibitem [{\citenamefont {Pikulin}\ \emph {et~al.}(2012)\citenamefont {Pikulin}, \citenamefont {Dahlhaus}, \citenamefont {Wimmer}, \citenamefont {Schomerus},\ and\ \citenamefont {Beenakker}}]{PikulinNJP2012}%
  \BibitemOpen
  \bibfield  {author} {\bibinfo {author} {\bibfnamefont {D.~I.}\ \bibnamefont {Pikulin}}, \bibinfo {author} {\bibfnamefont {J.~P.}\ \bibnamefont {Dahlhaus}}, \bibinfo {author} {\bibfnamefont {M.}~\bibnamefont {Wimmer}}, \bibinfo {author} {\bibfnamefont {H.}~\bibnamefont {Schomerus}}, \ and\ \bibinfo {author} {\bibfnamefont {C.~W.~J.}\ \bibnamefont {Beenakker}},\ }\href {http://stacks.iop.org/1367-2630/14/i=12/a=125011} {\bibfield  {journal} {\bibinfo  {journal} {New Journal of Physics}\ }\textbf {\bibinfo {volume} {14}},\ \bibinfo {pages} {125011} (\bibinfo {year} {2012})}\BibitemShut {NoStop}%
\bibitem [{\citenamefont {Popinciuc}\ \emph {et~al.}(2012)\citenamefont {Popinciuc}, \citenamefont {Calado}, \citenamefont {Liu}, \citenamefont {Akhmerov}, \citenamefont {Klapwijk},\ and\ \citenamefont {Vandersypen}}]{PopinciucPRB12}%
  \BibitemOpen
  \bibfield  {author} {\bibinfo {author} {\bibfnamefont {M.}~\bibnamefont {Popinciuc}}, \bibinfo {author} {\bibfnamefont {V.~E.}\ \bibnamefont {Calado}}, \bibinfo {author} {\bibfnamefont {X.~L.}\ \bibnamefont {Liu}}, \bibinfo {author} {\bibfnamefont {A.~R.}\ \bibnamefont {Akhmerov}}, \bibinfo {author} {\bibfnamefont {T.~M.}\ \bibnamefont {Klapwijk}}, \ and\ \bibinfo {author} {\bibfnamefont {L.~M.~K.}\ \bibnamefont {Vandersypen}},\ }\href {\doibase 10.1103/PhysRevB.85.205404} {\bibfield  {journal} {\bibinfo  {journal} {Phys. Rev. B}\ }\textbf {\bibinfo {volume} {85}},\ \bibinfo {pages} {205404} (\bibinfo {year} {2012})}\BibitemShut {NoStop}%
\bibitem [{\citenamefont {Zuo}\ \emph {et~al.}(2017)\citenamefont {Zuo}, \citenamefont {Mourik}, \citenamefont {Szombati}, \citenamefont {Nijholt}, \citenamefont {van Woerkom}, \citenamefont {Geresdi}, \citenamefont {Chen}, \citenamefont {Ostroukh}, \citenamefont {Akhmerov}, \citenamefont {Plissard}, \citenamefont {Car}, \citenamefont {Bakkers}, \citenamefont {Pikulin}, \citenamefont {Kouwenhoven},\ and\ \citenamefont {Frolov}}]{ZuoPRL17}%
  \BibitemOpen
  \bibfield  {author} {\bibinfo {author} {\bibfnamefont {K.}~\bibnamefont {Zuo}}, \bibinfo {author} {\bibfnamefont {V.}~\bibnamefont {Mourik}}, \bibinfo {author} {\bibfnamefont {D.~B.}\ \bibnamefont {Szombati}}, \bibinfo {author} {\bibfnamefont {B.}~\bibnamefont {Nijholt}}, \bibinfo {author} {\bibfnamefont {D.~J.}\ \bibnamefont {van Woerkom}}, \bibinfo {author} {\bibfnamefont {A.}~\bibnamefont {Geresdi}}, \bibinfo {author} {\bibfnamefont {J.}~\bibnamefont {Chen}}, \bibinfo {author} {\bibfnamefont {V.~P.}\ \bibnamefont {Ostroukh}}, \bibinfo {author} {\bibfnamefont {A.~R.}\ \bibnamefont {Akhmerov}}, \bibinfo {author} {\bibfnamefont {S.~R.}\ \bibnamefont {Plissard}}, \bibinfo {author} {\bibfnamefont {D.}~\bibnamefont {Car}}, \bibinfo {author} {\bibfnamefont {E.~P. A.~M.}\ \bibnamefont {Bakkers}}, \bibinfo {author} {\bibfnamefont {D.~I.}\ \bibnamefont {Pikulin}}, \bibinfo {author} {\bibfnamefont {L.~P.}\ \bibnamefont {Kouwenhoven}}, \ and\ \bibinfo {author} {\bibfnamefont {S.~M.}\ \bibnamefont {Frolov}},\ }\href
  {\doibase 10.1103/PhysRevLett.119.187704} {\bibfield  {journal} {\bibinfo  {journal} {Phys. Rev. Lett.}\ }\textbf {\bibinfo {volume} {119}},\ \bibinfo {pages} {187704} (\bibinfo {year} {2017})}\BibitemShut {NoStop}%
\bibitem [{\citenamefont {Lee}\ \emph {et~al.}(2014)\citenamefont {Lee}, \citenamefont {Jiang}, \citenamefont {Houzet}, \citenamefont {Aguado}, \citenamefont {Lieber},\ and\ \citenamefont {De~Franceschi}}]{LeeNatnano2014}%
  \BibitemOpen
  \bibfield  {author} {\bibinfo {author} {\bibfnamefont {E.~J.}\ \bibnamefont {Lee}}, \bibinfo {author} {\bibfnamefont {X.}~\bibnamefont {Jiang}}, \bibinfo {author} {\bibfnamefont {M.}~\bibnamefont {Houzet}}, \bibinfo {author} {\bibfnamefont {R.}~\bibnamefont {Aguado}}, \bibinfo {author} {\bibfnamefont {C.~M.}\ \bibnamefont {Lieber}}, \ and\ \bibinfo {author} {\bibfnamefont {S.}~\bibnamefont {De~Franceschi}},\ }\href@noop {} {\bibfield  {journal} {\bibinfo  {journal} {Nature nanotechnology}\ }\textbf {\bibinfo {volume} {9}},\ \bibinfo {pages} {79} (\bibinfo {year} {2014})}\BibitemShut {NoStop}%
\bibitem [{\citenamefont {Kells}\ \emph {et~al.}(2012)\citenamefont {Kells}, \citenamefont {Meidan},\ and\ \citenamefont {Brouwer}}]{KellsPRB2012}%
  \BibitemOpen
  \bibfield  {author} {\bibinfo {author} {\bibfnamefont {G.}~\bibnamefont {Kells}}, \bibinfo {author} {\bibfnamefont {D.}~\bibnamefont {Meidan}}, \ and\ \bibinfo {author} {\bibfnamefont {P.~W.}\ \bibnamefont {Brouwer}},\ }\href {\doibase 10.1103/PhysRevB.86.100503} {\bibfield  {journal} {\bibinfo  {journal} {Phys. Rev. B}\ }\textbf {\bibinfo {volume} {86}},\ \bibinfo {pages} {100503} (\bibinfo {year} {2012})}\BibitemShut {NoStop}%
\bibitem [{\citenamefont {Moore}\ \emph {et~al.}(2018)\citenamefont {Moore}, \citenamefont {Stanescu},\ and\ \citenamefont {Tewari}}]{Moore2018}%
  \BibitemOpen
  \bibfield  {author} {\bibinfo {author} {\bibfnamefont {C.}~\bibnamefont {Moore}}, \bibinfo {author} {\bibfnamefont {T.~D.}\ \bibnamefont {Stanescu}}, \ and\ \bibinfo {author} {\bibfnamefont {S.}~\bibnamefont {Tewari}},\ }\href {\doibase 10.1103/PhysRevB.97.165302} {\bibfield  {journal} {\bibinfo  {journal} {Phys. Rev. B}\ }\textbf {\bibinfo {volume} {97}},\ \bibinfo {pages} {165302} (\bibinfo {year} {2018})}\BibitemShut {NoStop}%
\bibitem [{\citenamefont {Vuik}\ \emph {et~al.}(2018)\citenamefont {Vuik}, \citenamefont {Nijholt}, \citenamefont {Akhmerov},\ and\ \citenamefont {Wimmer}}]{vuik2018}%
  \BibitemOpen
  \bibfield  {author} {\bibinfo {author} {\bibfnamefont {A.}~\bibnamefont {Vuik}}, \bibinfo {author} {\bibfnamefont {B.}~\bibnamefont {Nijholt}}, \bibinfo {author} {\bibfnamefont {A.}~\bibnamefont {Akhmerov}}, \ and\ \bibinfo {author} {\bibfnamefont {M.}~\bibnamefont {Wimmer}},\ }\href@noop {} {\bibfield  {journal} {\bibinfo  {journal} {arXiv preprint arXiv:1806.02801}\ } (\bibinfo {year} {2018})}\BibitemShut {NoStop}%
\bibitem [{\citenamefont {{Woods, Benjamin D. and Chen, Jun and Frolov, Sergey M., and Stanescu, Tudor D.}}(2019)}]{Woods2019}%
  \BibitemOpen
  \bibfield  {author} {\bibinfo {author} {\bibnamefont {{Woods, Benjamin D. and Chen, Jun and Frolov, Sergey M., and Stanescu, Tudor D.}}},\ }\href {https://arxiv.org/abs/1902.02772} {\bibfield  {journal} {\bibinfo  {journal} {arXiv:1902.02772}\ } (\bibinfo {year} {2019})}\BibitemShut {NoStop}%
\bibitem [{\citenamefont {Krogstrup}\ \emph {et~al.}(2015)\citenamefont {Krogstrup}, \citenamefont {Ziino}, \citenamefont {Chang}, \citenamefont {Albrecht}, \citenamefont {Madsen}, \citenamefont {Johnson}, \citenamefont {Nyg{\r a}rd}, \citenamefont {Marcus},\ and\ \citenamefont {Jespersen}}]{krogstrup2015}%
  \BibitemOpen
  \bibfield  {author} {\bibinfo {author} {\bibfnamefont {P.}~\bibnamefont {Krogstrup}}, \bibinfo {author} {\bibfnamefont {N.~L.~B.}\ \bibnamefont {Ziino}}, \bibinfo {author} {\bibfnamefont {W.}~\bibnamefont {Chang}}, \bibinfo {author} {\bibfnamefont {S.~M.}\ \bibnamefont {Albrecht}}, \bibinfo {author} {\bibfnamefont {M.~H.}\ \bibnamefont {Madsen}}, \bibinfo {author} {\bibfnamefont {E.}~\bibnamefont {Johnson}}, \bibinfo {author} {\bibfnamefont {J.}~\bibnamefont {Nyg{\r a}rd}}, \bibinfo {author} {\bibfnamefont {C.~M.}\ \bibnamefont {Marcus}}, \ and\ \bibinfo {author} {\bibfnamefont {T.~S.}\ \bibnamefont {Jespersen}},\ }\href {\doibase 10.1038/nmat4176} {\bibfield  {journal} {\bibinfo  {journal} {Nature Materials}\ }\textbf {\bibinfo {volume} {14}},\ \bibinfo {pages} {400} (\bibinfo {year} {2015})}\BibitemShut {NoStop}%
\bibitem [{\citenamefont {Shabani}\ \emph {et~al.}(2016)\citenamefont {Shabani}, \citenamefont {Kjaergaard}, \citenamefont {Suominen}, \citenamefont {Kim}, \citenamefont {Nichele}, \citenamefont {Pakrouski}, \citenamefont {Stankevic}, \citenamefont {Lutchyn}, \citenamefont {Krogstrup}, \citenamefont {Feidenhans'l}, \citenamefont {Kraemer}, \citenamefont {Nayak}, \citenamefont {Troyer}, \citenamefont {Marcus},\ and\ \citenamefont {Palmstr\o{}m}}]{Shabani2016}%
  \BibitemOpen
  \bibfield  {author} {\bibinfo {author} {\bibfnamefont {J.}~\bibnamefont {Shabani}}, \bibinfo {author} {\bibfnamefont {M.}~\bibnamefont {Kjaergaard}}, \bibinfo {author} {\bibfnamefont {H.~J.}\ \bibnamefont {Suominen}}, \bibinfo {author} {\bibfnamefont {Y.}~\bibnamefont {Kim}}, \bibinfo {author} {\bibfnamefont {F.}~\bibnamefont {Nichele}}, \bibinfo {author} {\bibfnamefont {K.}~\bibnamefont {Pakrouski}}, \bibinfo {author} {\bibfnamefont {T.}~\bibnamefont {Stankevic}}, \bibinfo {author} {\bibfnamefont {R.~M.}\ \bibnamefont {Lutchyn}}, \bibinfo {author} {\bibfnamefont {P.}~\bibnamefont {Krogstrup}}, \bibinfo {author} {\bibfnamefont {R.}~\bibnamefont {Feidenhans'l}}, \bibinfo {author} {\bibfnamefont {S.}~\bibnamefont {Kraemer}}, \bibinfo {author} {\bibfnamefont {C.}~\bibnamefont {Nayak}}, \bibinfo {author} {\bibfnamefont {M.}~\bibnamefont {Troyer}}, \bibinfo {author} {\bibfnamefont {C.~M.}\ \bibnamefont {Marcus}}, \ and\ \bibinfo {author} {\bibfnamefont {C.~J.}\ \bibnamefont {Palmstr\o{}m}},\ }\href {\doibase
  10.1103/PhysRevB.93.155402} {\bibfield  {journal} {\bibinfo  {journal} {Phys. Rev. B}\ }\textbf {\bibinfo {volume} {93}},\ \bibinfo {pages} {155402} (\bibinfo {year} {2016})}\BibitemShut {NoStop}%
\bibitem [{\citenamefont {Gazibegovic}\ \emph {et~al.}(2013)\citenamefont {Gazibegovic}, \citenamefont {Car}, \citenamefont {Zhang}, \citenamefont {Balk}, \citenamefont {Logan}, \citenamefont {de~Moor}, \citenamefont {Cassidy}, \citenamefont {Schmits}, \citenamefont {Xu}, \citenamefont {Wang}, \citenamefont {Krogstrup}, \citenamefont {het Veld}, \citenamefont {Zuo}, \citenamefont {Vos}, \citenamefont {Shen}, \citenamefont {Bouman}, \citenamefont {Shojaei}, \citenamefont {Pennachio}, \citenamefont {Lee}, \citenamefont {van Veldhoven}, \citenamefont {Koelling}, \citenamefont {Verheijen}, \citenamefont {Kouwenhoven}, \citenamefont {Palmstr{\o}m},\ and\ \citenamefont {Bakkers}}]{Gazibegovicnature2017}%
  \BibitemOpen
  \bibfield  {author} {\bibinfo {author} {\bibfnamefont {S.}~\bibnamefont {Gazibegovic}}, \bibinfo {author} {\bibfnamefont {D.}~\bibnamefont {Car}}, \bibinfo {author} {\bibfnamefont {H.}~\bibnamefont {Zhang}}, \bibinfo {author} {\bibfnamefont {S.~C.}\ \bibnamefont {Balk}}, \bibinfo {author} {\bibfnamefont {J.~A.}\ \bibnamefont {Logan}}, \bibinfo {author} {\bibfnamefont {M.~W.~A.}\ \bibnamefont {de~Moor}}, \bibinfo {author} {\bibfnamefont {M.~C.}\ \bibnamefont {Cassidy}}, \bibinfo {author} {\bibfnamefont {R.}~\bibnamefont {Schmits}}, \bibinfo {author} {\bibfnamefont {D.}~\bibnamefont {Xu}}, \bibinfo {author} {\bibfnamefont {G.}~\bibnamefont {Wang}}, \bibinfo {author} {\bibfnamefont {P.}~\bibnamefont {Krogstrup}}, \bibinfo {author} {\bibfnamefont {R.~L. M.~O.}\ \bibnamefont {het Veld}}, \bibinfo {author} {\bibfnamefont {K.}~\bibnamefont {Zuo}}, \bibinfo {author} {\bibfnamefont {Y.}~\bibnamefont {Vos}}, \bibinfo {author} {\bibfnamefont {J.}~\bibnamefont {Shen}}, \bibinfo {author} {\bibfnamefont {D.}~\bibnamefont
  {Bouman}}, \bibinfo {author} {\bibfnamefont {B.}~\bibnamefont {Shojaei}}, \bibinfo {author} {\bibfnamefont {D.}~\bibnamefont {Pennachio}}, \bibinfo {author} {\bibfnamefont {J.~S.}\ \bibnamefont {Lee}}, \bibinfo {author} {\bibfnamefont {P.~J.}\ \bibnamefont {van Veldhoven}}, \bibinfo {author} {\bibfnamefont {S.}~\bibnamefont {Koelling}}, \bibinfo {author} {\bibfnamefont {M.~A.}\ \bibnamefont {Verheijen}}, \bibinfo {author} {\bibfnamefont {L.~P.}\ \bibnamefont {Kouwenhoven}}, \bibinfo {author} {\bibfnamefont {C.~J.}\ \bibnamefont {Palmstr{\o}m}}, \ and\ \bibinfo {author} {\bibfnamefont {E.~P. A.~M.}\ \bibnamefont {Bakkers}},\ }\href {\doibase doi:10.1038/nature23468} {\bibfield  {journal} {\bibinfo  {journal} {Nature}\ }\textbf {\bibinfo {volume} {548}},\ \bibinfo {pages} {434} (\bibinfo {year} {2013})}\BibitemShut {NoStop}%
\bibitem [{\citenamefont {Albrecht}\ \emph {et~al.}(2016{\natexlab{b}})\citenamefont {Albrecht}, \citenamefont {Higginbotham}, \citenamefont {Madsen}, \citenamefont {Kuemmeth}, \citenamefont {Jespersen}, \citenamefont {Nyg{\aa}rd}, \citenamefont {Krogstrup},\ and\ \citenamefont {Marcus}}]{AlbrechtNature2016}%
  \BibitemOpen
  \bibfield  {author} {\bibinfo {author} {\bibfnamefont {S.~M.}\ \bibnamefont {Albrecht}}, \bibinfo {author} {\bibfnamefont {A.~P.}\ \bibnamefont {Higginbotham}}, \bibinfo {author} {\bibfnamefont {M.}~\bibnamefont {Madsen}}, \bibinfo {author} {\bibfnamefont {F.}~\bibnamefont {Kuemmeth}}, \bibinfo {author} {\bibfnamefont {T.~S.}\ \bibnamefont {Jespersen}}, \bibinfo {author} {\bibfnamefont {J.}~\bibnamefont {Nyg{\aa}rd}}, \bibinfo {author} {\bibfnamefont {P.}~\bibnamefont {Krogstrup}}, \ and\ \bibinfo {author} {\bibfnamefont {C.~M.}\ \bibnamefont {Marcus}},\ }\href@noop {} {\bibfield  {journal} {\bibinfo  {journal} {Nature}\ }\textbf {\bibinfo {volume} {531}},\ \bibinfo {pages} {206} (\bibinfo {year} {2016}{\natexlab{b}})}\BibitemShut {NoStop}%
\bibitem [{\citenamefont {Vuik}\ \emph {et~al.}(2016)\citenamefont {Vuik}, \citenamefont {Eeltink}, \citenamefont {Akhmerov},\ and\ \citenamefont {Wimmer}}]{VuikNJP2016}%
  \BibitemOpen
  \bibfield  {author} {\bibinfo {author} {\bibfnamefont {A.}~\bibnamefont {Vuik}}, \bibinfo {author} {\bibfnamefont {D.}~\bibnamefont {Eeltink}}, \bibinfo {author} {\bibfnamefont {A.}~\bibnamefont {Akhmerov}}, \ and\ \bibinfo {author} {\bibfnamefont {M.}~\bibnamefont {Wimmer}},\ }\href@noop {} {\bibfield  {journal} {\bibinfo  {journal} {New Journal of Physics}\ }\textbf {\bibinfo {volume} {18}},\ \bibinfo {pages} {033013} (\bibinfo {year} {2016})}\BibitemShut {NoStop}%
\bibitem [{\citenamefont {Stanescu}\ \emph {et~al.}(2013)\citenamefont {Stanescu}, \citenamefont {Lutchyn},\ and\ \citenamefont {Sarma}}]{StanescuPRB2013}%
  \BibitemOpen
  \bibfield  {author} {\bibinfo {author} {\bibfnamefont {T.~D.}\ \bibnamefont {Stanescu}}, \bibinfo {author} {\bibfnamefont {R.~M.}\ \bibnamefont {Lutchyn}}, \ and\ \bibinfo {author} {\bibfnamefont {S.~D.}\ \bibnamefont {Sarma}},\ }\href@noop {} {\bibfield  {journal} {\bibinfo  {journal} {Physical Review B}\ }\textbf {\bibinfo {volume} {87}},\ \bibinfo {pages} {094518} (\bibinfo {year} {2013})}\BibitemShut {NoStop}%
\bibitem [{\citenamefont {van Weperen}\ \emph {et~al.}(2013)\citenamefont {van Weperen}, \citenamefont {Plissard}, \citenamefont {Bakkers}, \citenamefont {Frolov},\ and\ \citenamefont {Kouwenhoven}}]{vanweperennanolett13}%
  \BibitemOpen
  \bibfield  {author} {\bibinfo {author} {\bibfnamefont {I.}~\bibnamefont {van Weperen}}, \bibinfo {author} {\bibfnamefont {S.~R.}\ \bibnamefont {Plissard}}, \bibinfo {author} {\bibfnamefont {E.~P. A.~M.}\ \bibnamefont {Bakkers}}, \bibinfo {author} {\bibfnamefont {S.~M.}\ \bibnamefont {Frolov}}, \ and\ \bibinfo {author} {\bibfnamefont {L.~P.}\ \bibnamefont {Kouwenhoven}},\ }\href@noop {} {\bibfield  {journal} {\bibinfo  {journal} {Nano Letters}\ }\textbf {\bibinfo {volume} {13}},\ \bibinfo {pages} {387} (\bibinfo {year} {2013})}\BibitemShut {NoStop}%
\bibitem [{\citenamefont {Eichler}\ \emph {et~al.}(2007)\citenamefont {Eichler}, \citenamefont {Weiss}, \citenamefont {Oberholzer}, \citenamefont {Sch{\"o}nenberger}, \citenamefont {Yeyati}, \citenamefont {Cuevas},\ and\ \citenamefont {Mart{\'\i}n-Rodero}}]{eichlerprl2007}%
  \BibitemOpen
  \bibfield  {author} {\bibinfo {author} {\bibfnamefont {A.}~\bibnamefont {Eichler}}, \bibinfo {author} {\bibfnamefont {M.}~\bibnamefont {Weiss}}, \bibinfo {author} {\bibfnamefont {S.}~\bibnamefont {Oberholzer}}, \bibinfo {author} {\bibfnamefont {C.}~\bibnamefont {Sch{\"o}nenberger}}, \bibinfo {author} {\bibfnamefont {A.~L.}\ \bibnamefont {Yeyati}}, \bibinfo {author} {\bibfnamefont {J.}~\bibnamefont {Cuevas}}, \ and\ \bibinfo {author} {\bibfnamefont {A.}~\bibnamefont {Mart{\'\i}n-Rodero}},\ }\href@noop {} {\bibfield  {journal} {\bibinfo  {journal} {Physical review letters}\ }\textbf {\bibinfo {volume} {99}},\ \bibinfo {pages} {126602} (\bibinfo {year} {2007})}\BibitemShut {NoStop}%
\bibitem [{\citenamefont {Woods}\ \emph {et~al.}(2018)\citenamefont {Woods}, \citenamefont {Stanescu},\ and\ \citenamefont {Das~Sarma}}]{Woods2018a}%
  \BibitemOpen
  \bibfield  {author} {\bibinfo {author} {\bibfnamefont {B.~D.}\ \bibnamefont {Woods}}, \bibinfo {author} {\bibfnamefont {T.~D.}\ \bibnamefont {Stanescu}}, \ and\ \bibinfo {author} {\bibfnamefont {S.}~\bibnamefont {Das~Sarma}},\ }\href {\doibase 10.1103/PhysRevB.98.035428} {\bibfield  {journal} {\bibinfo  {journal} {Phys. Rev. B}\ }\textbf {\bibinfo {volume} {98}},\ \bibinfo {pages} {035428} (\bibinfo {year} {2018})}\BibitemShut {NoStop}%
\bibitem [{\citenamefont {Blonder}\ \emph {et~al.}(1982)\citenamefont {Blonder}, \citenamefont {Tinkham},\ and\ \citenamefont {Klapwijk}}]{BlonderPhysRevB1982}%
  \BibitemOpen
  \bibfield  {author} {\bibinfo {author} {\bibfnamefont {G.}~\bibnamefont {Blonder}}, \bibinfo {author} {\bibfnamefont {M.}~\bibnamefont {Tinkham}}, \ and\ \bibinfo {author} {\bibfnamefont {T.}~\bibnamefont {Klapwijk}},\ }\href@noop {} {\bibfield  {journal} {\bibinfo  {journal} {Physical Review B}\ }\textbf {\bibinfo {volume} {25}},\ \bibinfo {pages} {4515} (\bibinfo {year} {1982})}\BibitemShut {NoStop}%
\bibitem [{\citenamefont {Rokhinson}\ \emph {et~al.}(2012)\citenamefont {Rokhinson}, \citenamefont {Liu},\ and\ \citenamefont {Furdyna}}]{RokhinsonNPhys2012}%
  \BibitemOpen
  \bibfield  {author} {\bibinfo {author} {\bibfnamefont {L.~P.}\ \bibnamefont {Rokhinson}}, \bibinfo {author} {\bibfnamefont {X.}~\bibnamefont {Liu}}, \ and\ \bibinfo {author} {\bibfnamefont {J.~K.}\ \bibnamefont {Furdyna}},\ }\href {\doibase 10.1038/nphys2429} {\bibfield  {journal} {\bibinfo  {journal} {Nature Physics}\ }\textbf {\bibinfo {volume} {8}},\ \bibinfo {pages} {795} (\bibinfo {year} {2012})}\BibitemShut {NoStop}%
\bibitem [{\citenamefont {Aasen}\ \emph {et~al.}(2016)\citenamefont {Aasen}, \citenamefont {Hell}, \citenamefont {Mishmash}, \citenamefont {Higginbotham}, \citenamefont {Danon}, \citenamefont {Leijnse}, \citenamefont {Jespersen}, \citenamefont {Folk}, \citenamefont {Marcus}, \citenamefont {Flensberg},\ and\ \citenamefont {Alicea}}]{AasenPRX2016}%
  \BibitemOpen
  \bibfield  {author} {\bibinfo {author} {\bibfnamefont {D.}~\bibnamefont {Aasen}}, \bibinfo {author} {\bibfnamefont {M.}~\bibnamefont {Hell}}, \bibinfo {author} {\bibfnamefont {R.~V.}\ \bibnamefont {Mishmash}}, \bibinfo {author} {\bibfnamefont {A.}~\bibnamefont {Higginbotham}}, \bibinfo {author} {\bibfnamefont {J.}~\bibnamefont {Danon}}, \bibinfo {author} {\bibfnamefont {M.}~\bibnamefont {Leijnse}}, \bibinfo {author} {\bibfnamefont {T.~S.}\ \bibnamefont {Jespersen}}, \bibinfo {author} {\bibfnamefont {J.~A.}\ \bibnamefont {Folk}}, \bibinfo {author} {\bibfnamefont {C.~M.}\ \bibnamefont {Marcus}}, \bibinfo {author} {\bibfnamefont {K.}~\bibnamefont {Flensberg}}, \ and\ \bibinfo {author} {\bibfnamefont {J.}~\bibnamefont {Alicea}},\ }\href {\doibase 10.1103/PhysRevX.6.031016} {\bibfield  {journal} {\bibinfo  {journal} {Phys. Rev. X}\ }\textbf {\bibinfo {volume} {6}},\ \bibinfo {pages} {031016} (\bibinfo {year} {2016})}\BibitemShut {NoStop}%
\bibitem [{\citenamefont {Houzet}\ \emph {et~al.}(2013)\citenamefont {Houzet}, \citenamefont {Meyer}, \citenamefont {Badiane},\ and\ \citenamefont {Glazman}}]{houzetprl2013}%
  \BibitemOpen
  \bibfield  {author} {\bibinfo {author} {\bibfnamefont {M.}~\bibnamefont {Houzet}}, \bibinfo {author} {\bibfnamefont {J.~S.}\ \bibnamefont {Meyer}}, \bibinfo {author} {\bibfnamefont {D.~M.}\ \bibnamefont {Badiane}}, \ and\ \bibinfo {author} {\bibfnamefont {L.~I.}\ \bibnamefont {Glazman}},\ }\href@noop {} {\bibfield  {journal} {\bibinfo  {journal} {Physical review letters}\ }\textbf {\bibinfo {volume} {111}},\ \bibinfo {pages} {046401} (\bibinfo {year} {2013})}\BibitemShut {NoStop}%
\bibitem [{\citenamefont {Vaitiek\ifmmode~\dot{e}\else \.{e}\fi{}nas}\ \emph {et~al.}(2018)\citenamefont {Vaitiek\ifmmode~\dot{e}\else \.{e}\fi{}nas}, \citenamefont {Deng}, \citenamefont {Nyg\aa{}rd}, \citenamefont {Krogstrup},\ and\ \citenamefont {Marcus}}]{Vaitiek2018PRL}%
  \BibitemOpen
  \bibfield  {author} {\bibinfo {author} {\bibfnamefont {S.}~\bibnamefont {Vaitiek\ifmmode~\dot{e}\else \.{e}\fi{}nas}}, \bibinfo {author} {\bibfnamefont {M.-T.}\ \bibnamefont {Deng}}, \bibinfo {author} {\bibfnamefont {J.}~\bibnamefont {Nyg\aa{}rd}}, \bibinfo {author} {\bibfnamefont {P.}~\bibnamefont {Krogstrup}}, \ and\ \bibinfo {author} {\bibfnamefont {C.~M.}\ \bibnamefont {Marcus}},\ }\href {\doibase 10.1103/PhysRevLett.121.037703} {\bibfield  {journal} {\bibinfo  {journal} {Phys. Rev. Lett.}\ }\textbf {\bibinfo {volume} {121}},\ \bibinfo {pages} {037703} (\bibinfo {year} {2018})}\BibitemShut {NoStop}%
\end{thebibliography}%

\clearpage

\newpage

\setcounter{figure}{0}
\setcounter{equation}{0}
\setcounter{section}{0}
\begin{widetext}

\section{Supplemental Material for 'Ubiquitous non-Majorana Zero-Bias Conductance Peaks in Nanowire Devices'}


\renewcommand{\thefigure}{S\arabic{figure}}
\setcounter{figure}{0}

\subsection{Experimental measurements conditions}

Measurements are performed in a dilution refrigerator at a base temperature of $30$ mK, by standard low-frequency lock-in technique ($77.77$ Hz, $5~\mu$V, time constant $100$ ms, setting time $300$ ms). Multiple stages of filtering are used to enhance signal-to-noise ratio(a $\pi$-filter at room temperature, a copper-powder filter and a RC filter at base temperature). Estimated electronic temperature is around $80$ mK, which is extracted from the minimal observed subgap peaks FWHM. DC bias is scanned in steps of $5-10~\mu$V. The voltage drop on the devices is calibrated (voltage drop on filters is subtracted). For all the measurements, bias voltage is applied to the normal contact and the superconducting contact is grounded. All the measurement data are plotted using Spyview with an image processing of lowpass $(1, 0)$, which smoothen the data in the bias voltage but not in gate or magnetic field.

\subsection{Fig. \ref{figS1_new} Line traces of conductance maps in the main text \textcolor{red}{and Devices configuration} }

\begin{figure*}[h!]
\centering
  \includegraphics[width=0.95\textwidth]{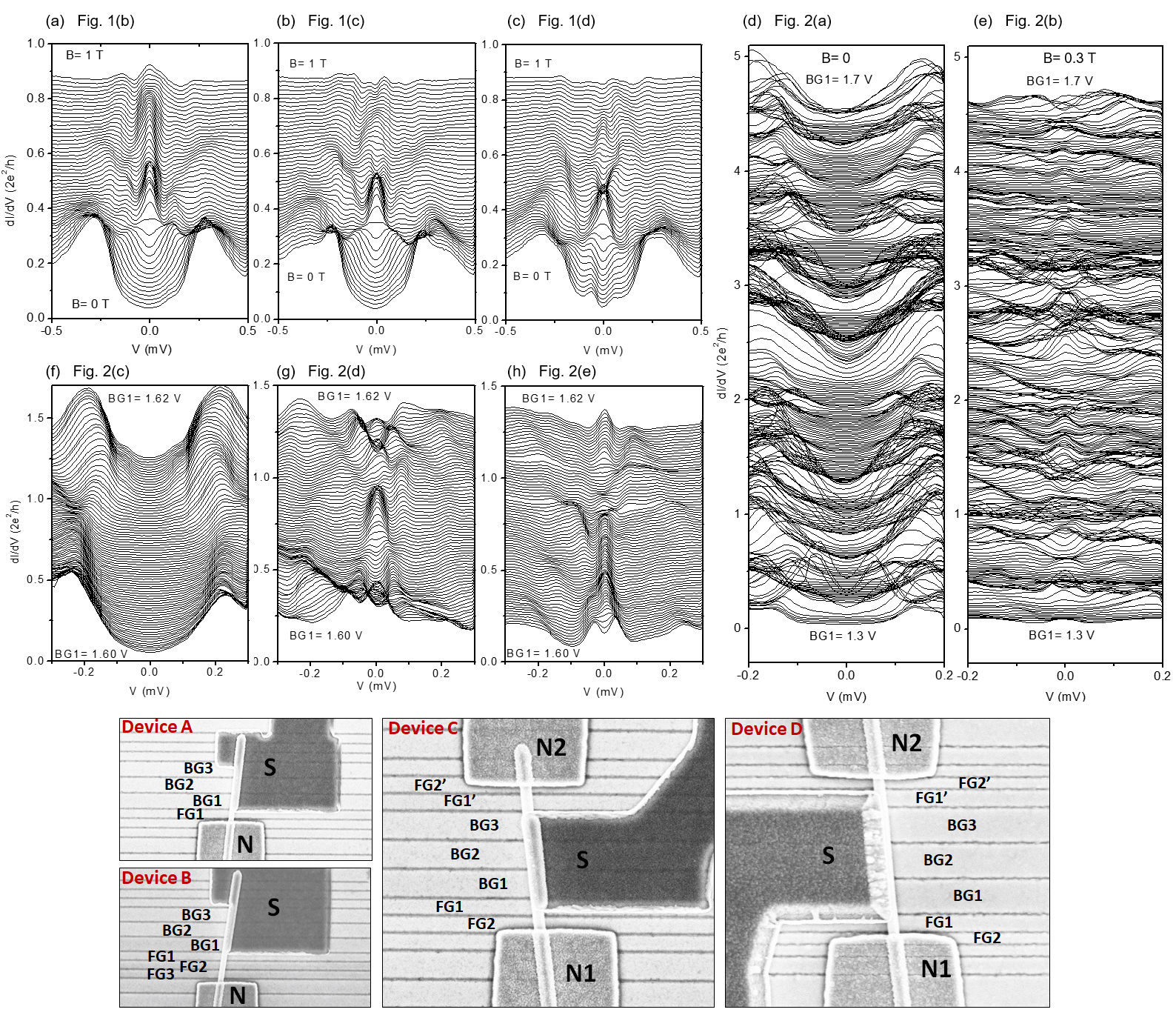}
\caption{
In this figure, we plot all the line traces of conductance maps in the main text, to help one understand the ZBP feature better. In all the panels, there is an offset of 0.015 (in the unit of $2e^2/h$) for clarify. The observed ZBP are dramatically longer than that expected from the overlapping of crossed states. Take an example in Fig. S1(a)(Fig. 1b in the main text), the observed ZBP extends for at least $0.5$ T in magnetic field. From the slope($0.7$ mV/T) and peak width(FWHM = $90~\mu$V) of the crossed states, we calculate the length of expected ZBP from crossed states overlapping, which is $0.128$ T, much smaller than the observed ZBP. \textcolor{red}{There were totally four devices measured in this paper. The device names are: 
Device A: 0520-840-NW5; Device B: 0520-840-2; Device C: 0520-840-NW1; Device D: 0510-197. The device configurations are listed here and the full data for all the devices is in Zenodo:https://doi.org/10.5281/zenodo.5793360}
 \label{figS1_new}}
\end{figure*}
\clearpage

\subsection{Fig. \ref{figS1} Zero bias resonances at magnetic field up to 2T}

\begin{figure*}[h!]
\centering
  \includegraphics[width=0.5\textwidth]{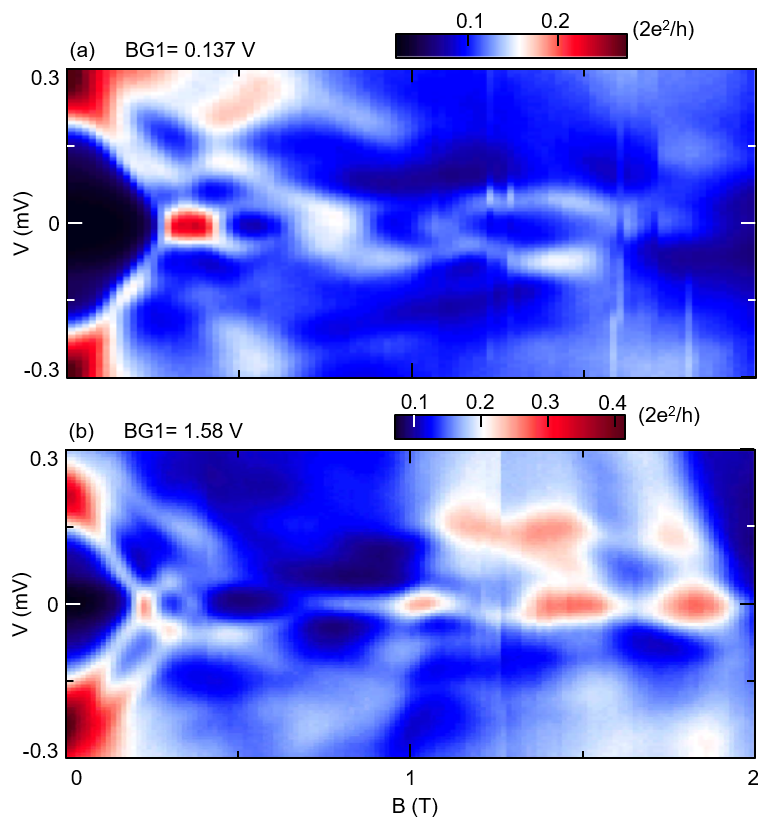}
\caption{
In this figure we show that zero bias peaks and near-zero oscillations can persist up to $2$~T. Panels (a) and (b) are differential conductance maps of bias voltage $V$ versus magnetic field for two very different $BG1$ settings. The device studied is the same as in the main text Fig. 1(a). In both (a) and (b), we see a gap closing around 0.3 T. In Fig. (a), the gap closes and generates a short zero-bias peak. Then the resonance splits and merges back to zero. In (b), a peak near zero bias persists from $0.7$~T to $2$~T.  In addition to the pinned zero-bias peak, we also see multiple branches of conductance resonances at high bias that move down to zero bias at higher fields, where they enhance the zero-bias resonance.
 \label{figS1}}
\end{figure*}
\clearpage

\subsection{Fig. \ref{figS2} Persistent ZBP in \textcolor{red}{Device C}}

\begin{figure*}[h!]
\centering
  \includegraphics[width=0.5\textwidth]{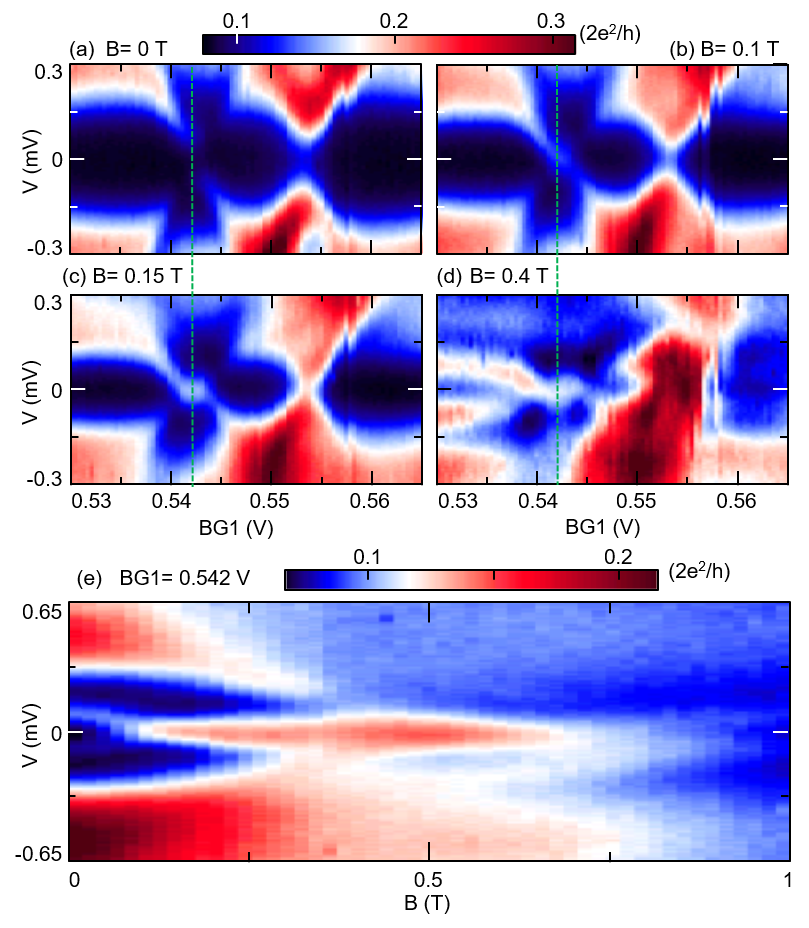}
\caption{
This figure discusses zero-bias resonance observed in \textcolor{red}{Device C}. (a)-(d) Differential conductance maps of bias voltage versus gate $BG1$ at different magnetic fields. At zero field, there are no pinned zero-bias resonances, while a weak transient conductance resonance is present at $BG1= 0.5402$~V. With magnetic field applied and increasing, a zero-bias peak appears at $BG1= 0.5402$~V (marked by green dashed lines) and persists above $0.4$~T. (e) At this particular setting of $BG1$, a scan of bias voltage versus magnetic field shows a zero-bias peak persisting up to $1$~T, having come from two mid-gap resonances at zero field. It is noteworthy that the induced gap seems to close and reopen around $0.4$~T. This long-lived zero-bias peak is only observed at one particular gate setting of $BG1= 0.5402$~V, while no pinned zero-bias peaks are observed upon minor variations in $BG1$.
 \label{figS2}}
\end{figure*}
\clearpage
\subsection{Fig. \ref{figS3} Study of ABS and their apparent g-factors in \textcolor{red}{Device A}}

 In Fig. \ref{figS3} we examine ABS in \textcolor{red}{Device A}. The apparent g-factor extracted from the dispersion of ABS in magnetic field can vary strongly with only minor variations in gate voltage. This is due to quantum dot singlet-doublet physics: in the even-parity regime the resonance is pinned to the gap and moves above the gap at higher field, leaving the gap edge to be the lowest energy resonance. In the odd-parity regime the resonance exhibits opposite behavior: it moves towards zero bias in magnetic field. At the transition between the even- and odd- parity regimes the competition of two opposite field behaviors causes the apparent g-factor to be strongly reduced.

\begin{figure*}[h!]
\centering
  \includegraphics[width=\textwidth]{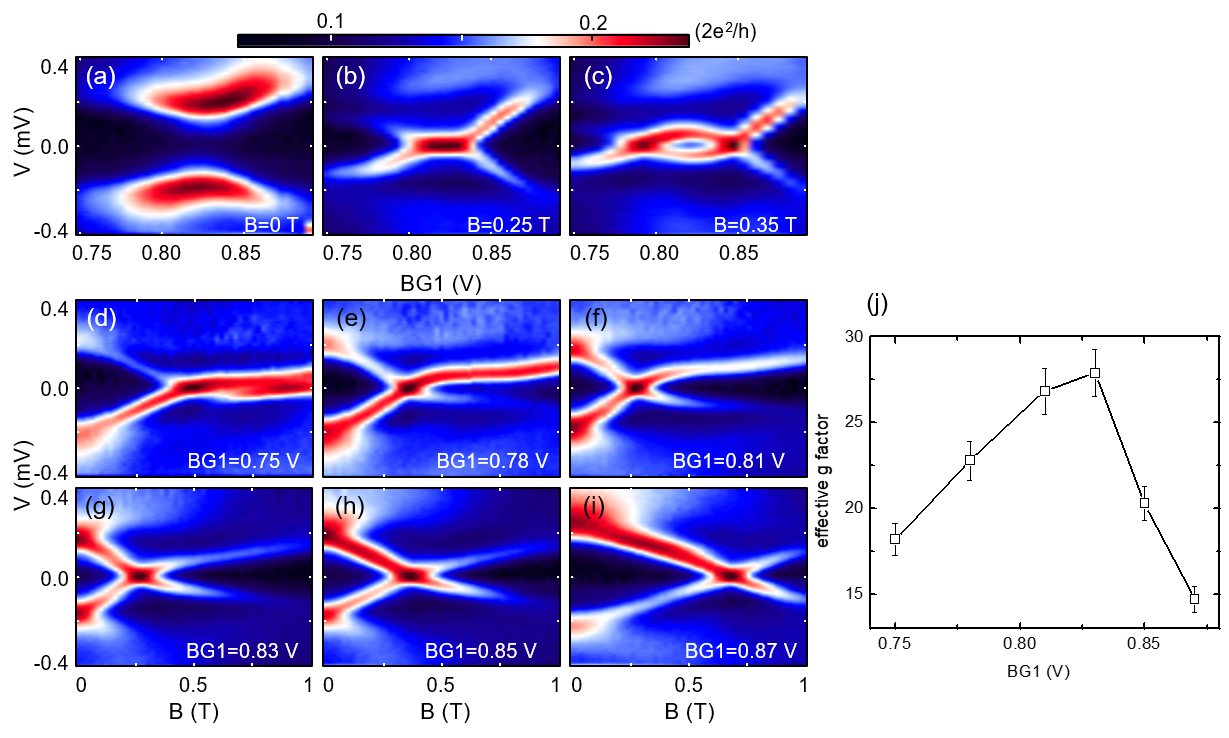}
\caption{
 (a)−(c) The evolution of ABS with magnetic field. (a) At zero field, a pair of conductance resonances appears symmetrically at positive and negative bias close to the superconducting gap edge. (b) At a finite field of $0.25$~T, conductance resonances split off from the gap edge and move to lower bias touching zero bias in the center of the scan. (c) At $0.35$~T top and bottom conductance resonances trade positions forming a loop in the center. This behavior is characteristic of a magnetic-field induced singlet-to-doublet ground state quantum phase transition in quantum dots coupled to superconductors \cite{LeeNatnano2014}. Panels (d)-(i) demonstrate how magnetic field dispersion of ABS is affected by $BG1$ set to different points throughout the range shown in panels (a)-(c). In panel (d), a near-zero resonance is observed over a significant range of magnetic field exceeding 0.5T. However, subsequent panels show that this pinning to zero bias is a result of fine-tuning and is only seen at a particular setting of $BG1$. The dispersion in magnetic field cannot be described by a straight line, because an inflection point is observed in all magnetic-field dependences. One possible explanation for this is level repulsion from other states, though those states are not directly visible as resonances in the data. With a slight change of gate $BG1$, the zero bias crossing point shifts by 100's of mT, while the energy of the resonance shifts only weakly at zero field, remaining in the 200-250$\mu$eV range (see panel (a). For this reason, effective g-factors plotted in panel  (j) show a large variation with minor changes in $BG1$. The g-factors are extracted by fitting the resonances to a straight line between zero field and the zero-bias crossing point. Significant g-factor tuning has been reported previously in nanowire devices \cite{Vaitiek2018PRL}. Here we show that it originates from the interplay of singlet and doublet physics in a evenly or oddly occupied quantum dot. 
 \label{figS3}}
\end{figure*}
\clearpage

\subsection{Fig. \ref{figS4} Two independent ABS in \textcolor{red}{Device A}}

\begin{figure*}[h!]
\centering
  \includegraphics[width=0.6\textwidth]{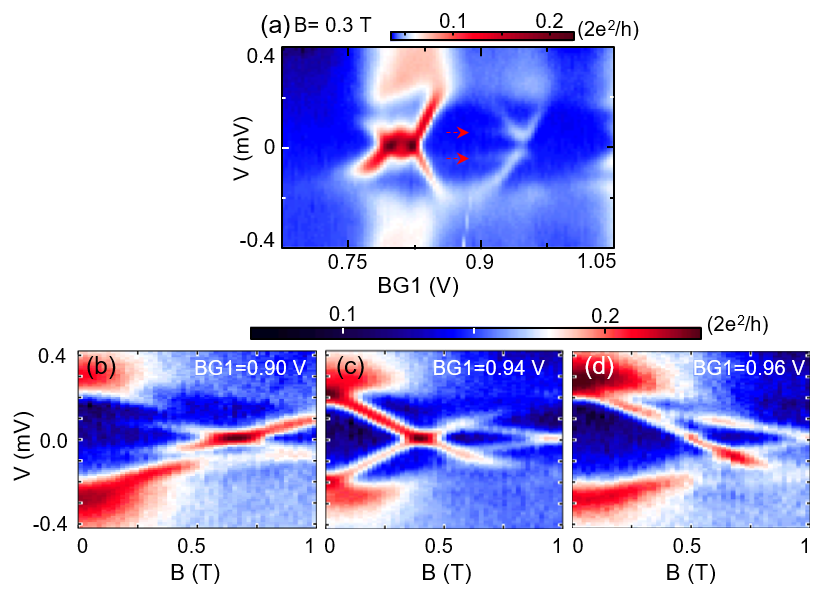}
\caption{
This figure is an extension of Fig.  \ref{figS3}, which demonstrates two sets of resonances dispersing in $BG1$ and $B$ in a close parameter range. Panel (a) shows two loop-like resonances at finite field. The left loop matches that in Fig.\ref{figS3}(c). The right loop does not yet touch zero bias, indicating a singlet-doublet quantum phase transition at a higher field. In (b)-(d), we see two sets of conductance resonances crossing zero bias at different magnetic fields. Notice that when the resonances from two sets cross they don't exhibit visible level repulsion, suggesting that these ABS do not interact but simply coexist. Also notice that resonance that yields a zero-bias crossing in panel (b) does not originate from the left loop, but rather from a set of weakly visible parallel lines(marked by dashed arrows).
 \label{figS4}}
\end{figure*}
\clearpage

\subsection{Fig. \ref{figS5} Gate-tunable ABS in \textcolor{red}{Device D}}

\begin{figure*}[h!]
\centering
  \includegraphics[width=0.9\textwidth]{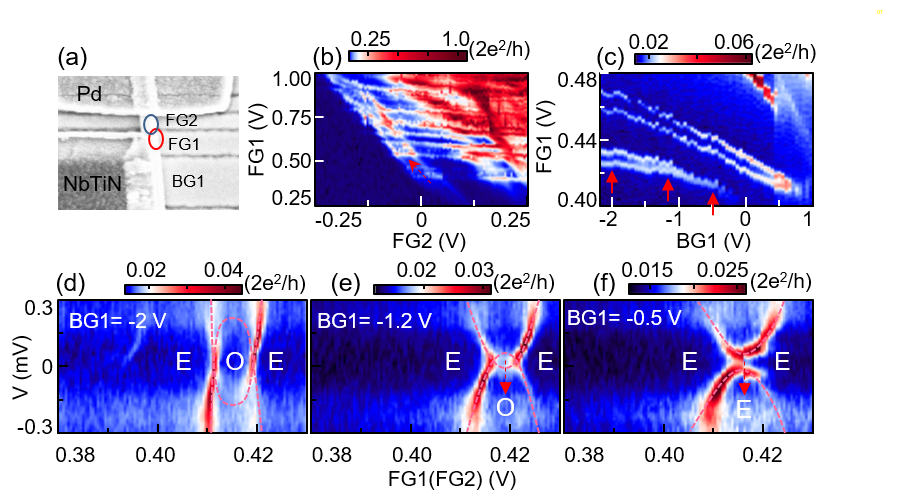}
\caption{(a) SEM image of \textcolor{red}{Device D}. The fabrication procedure and recipe are the same as the device in the main text. Blue and red ovals mark tentative locations for quantum dots studied in panel (b). Note that the gate layout is somewhat different for the third device, with two narrower gates $FG1$ and $FG2$ followed by a wider gate $BG1$. (b) Scan of $FG1$ versus $FG2$ at zero source-drain bias and zero external magnetic field. We observe two sets of conductance resonances, one set mostly horizontal, i.e. strongly tunable by $FG1$, while the other set tunable by both $FG1$ and $FG2$. This is consistent with the formation of two quantum dots in the $FG1-FG2$ segment of the device. Note also the similarity to Fig.4(a) in the main text where three families of resonances are observed. (c) Zero bias conductance map with $FG1, FG2$ tuned along the red arrow in panel (b), versus $BG1$. The conductance resonances (referring to loop structures in (d-f)) correspond to singlet-doublet ground state transitions, indicating that $BG1$ tunes the coupling between a quantum dot and a superconductor. At the most negative $BG1$, two pairs of conductance peaks show up along $FG1$($FG2$) axis. with $BG1$ increasing, each pair of peaks merge into one peak and vanish. At the merge point the quantum dot transitions into a strongly coupled regime with a BCS singlet ground state\cite{LeeNatnano2014}. Panels (d)-(f) confirm this through scans of bias versus $FG1$($FG2$) at three different settings of $BG1$ marked by arrows in panel (c). The dashed lines are guides to the eye. 
 \label{figS5}}
\end{figure*}
\clearpage

\subsection{Device setup in 3D simulations and additional simulation data}
\begin{figure*}[h!]
\centering
  \includegraphics[width=0.4\textwidth]{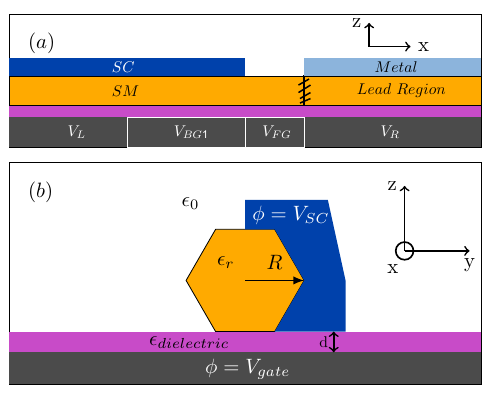}
\caption{(a) The semiconductor nanowire (orange, SM) is proximity coupled to a superconductor (blue, SC) and a metal (light blue) is the lead region. A dielectric layer (purple) separates the nanowire from potential gates (dark grey). There are four regions defined by the external gates and the materials deposited on the SM wire: the left bulk region  (gate potential $V_L$),  the big-gate region ($V_{BG1}$, $200$ nm), the fine-gate (uncovered) region ($V_{FG}$, $100$ nm), and the lead region ($V_R$). We apply a large negative voltage $V_L$ to the bulk gate, such that the low-energy states are localized primarily in the big-gate and fine-gate regions. The lead region is a continuation of the SM nanowire but in proximity to a metal, which strongly renormalizes its properties. We treat the coupling between the SM and lead region as a weak link (as indicated by black lines between the SM and the lead region). (b) Schematic representation of the cross section of the nanowire device in the SC-covered region. The superconductor is not explicitly included in the device Hamiltonian, but rather treated as a boundary condition in the electrostatics calculation.
 \label{figS6}}
\end{figure*}

\begin{figure*}[b!]
\centering
  \includegraphics[width=0.6\textwidth]{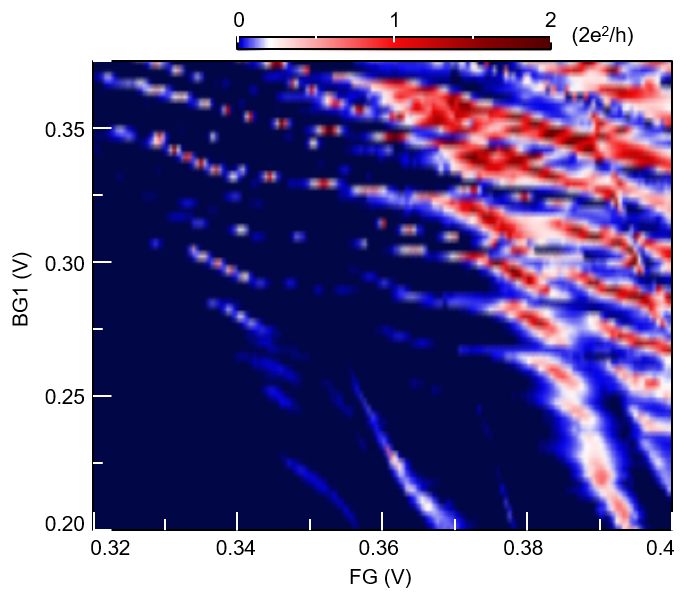}
\caption{Zero bias differential conductance map with reduced coupling between $FG$ and $BG1$ regions.
Hopping parameter between last and first layers of the $BG1$ and $FG$ regions is given by $t^{'} =0.7 t$, where $t$ is the hopping parameter between all other layers. 
 \label{figS7}}
\end{figure*}

A comparison of the experimental and theoretical zero bias differential conductance was shown in  Fig. 4 of the main text. While both conductance maps show peaks with three different slopes, the vertical (white) lines in the simulation are not as steep as those seen in the experiment. Indeed the vertical (white) lines in the experiments are nearly completely independent of the $BG1$ voltage, suggesting the $FG$ and $BG1$ regions are not as well coupled as our model predicts. Moreover, the features associated with the experimental vertical lines in Fig. 4(a) are broader than those seen in the simulation.

To study how reducing the coupling between the $FG$ and $BG1$ regions changes the conductance maps we reduce the hopping between the last and first layers of the $BG1$ and $FG$ regions, and recalculate the zero bias differential conductance. The result is shown in Fig. \ref{figS7} for $t^{'}=0.7t$, where $t^{'}$ is the hopping parameter between the last and first layers of the $BG1$ and $FG$ regions, respectively, and $t$ is the hopping parameter between all other layers. The conductance map has a broader and more vertical feature starting in the lower right corner that shows a closer resemblance to the vertical features in Fig. 4(a).

A possible explanation for the reduction in coupling involves our neglection of self-consistency in the charge distribution along the length of the wire. As mentioned in the main text and explained completely in Ref. \cite{Woods2018a}, our numerical method in general involves self-consistency both when constructing the effective 1D multi-orbital model and when solving the 1D problem. However, here we chose to ignore the self-consistency when solving the 1D problem due to the large parameter space that needed to be explored. This amounts to ignoring the charge redistribution along the length of the wire, which can produce potential barriers between gate regions \cite{Woods2018a}. This extra potential barrier between the $FG$ and $BG1$ regions would then explain the reduced coupling and the broader vertical features seen in the experiment.

\begin{figure*}[h!]
\centering
  \includegraphics[width=0.8\textwidth]{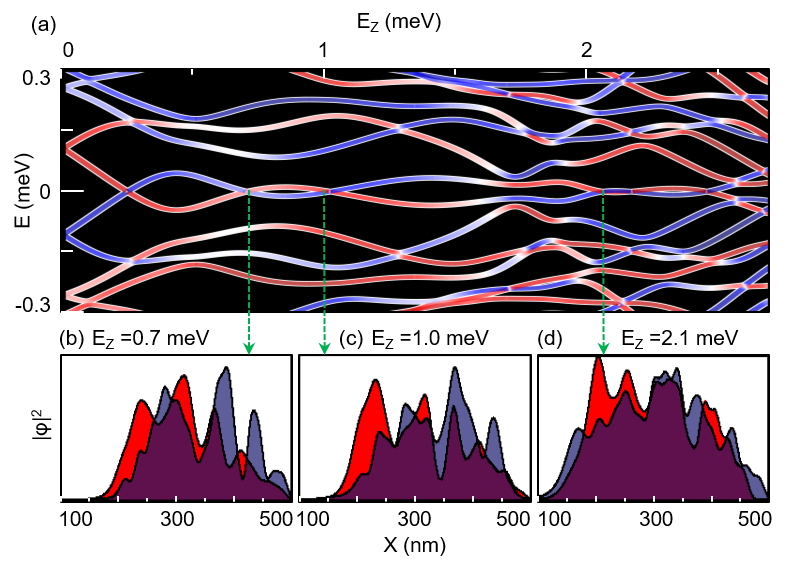}
\caption{(a) Spectrum of system without the lead attached. Gate parameters are given by $FG = 0.38$~V, $BG=0.353$~V to match those in Fig. 3(d). Note that the absence of lead slightly alters positions and shapes of features when compared to Fig. 3(d). (b)-(d) Majorana representation wavefunctions of zero energy states in (a) with dashed arrows indicating correspondence between the plots. Majorana wavefunctions are shown in red and blue filled curves, respectively. Large overlap between Majorana wavefunctions implies states are non-topological low-energy states.
 \label{figS8}}
\end{figure*}

\subsection{Majorana Representation of trivial low energy states}
We claim in the main text that zero bias peaks in the differential conductance scans arise from topological trivial states rather than topological MBSs. To quantify this claim, we make use of the Majorana representation of the zero energy states that are responsible for the zero bias peaks. Recall that any Bogoliubov-de Gennes (BdG) Hamiltonian can be written as a linear combination of two Majorana modes. Let $\phi_\epsilon$ and $\phi_{-\epsilon}$ be eigenstates of the BdG Hamiltonian that are related to each other by particle-hole symmetry with energies $\epsilon$ and $-\epsilon$, respectively. From these particle-hole symmetric states we can construct
\begin{align}
\psi_A = \frac{1}{\sqrt{2}} \left[\phi_\epsilon + \phi_{-\epsilon}\right], \\
\psi_B = \frac{i}{\sqrt{2}} \left[\phi_\epsilon - \phi_{-\epsilon}\right].
\end{align}
By construction these states satisfy the Majorana condition of being equal parts particle and hole, however they are not eigenstates of the BdG Hamiltonian unless $\epsilon=0$. Note that $\phi_{\pm \epsilon}=\frac{1}{\sqrt{2}}\left[\psi_A \pm i\psi_B\right]$. This is therefore named the Majorana representation of an eigenstate of the BdG Hamiltonian. For topological Majorana states, the wavefunctions $\psi_A$ and $\psi_B$ should be localized on opposite ends of the wire. In contrast, a trivial Andreev Bound state will have highly overlapping $\psi_A$ and $\psi_B$. Fig. \ref{figS8}(a) shows the spectrum of the setup without the lead attached. The parameters are the same as the differential conductance scan shown in Fig. 3(d). Note that the absence/presence of the lead alters the position and shapes of the features when comparing the spectrum to the differential conductance. We plot the Majorana wavefunctions in Figs. \ref{figS8}(b)-(d) coming from the corresponding zero energy states in Fig. \ref{figS8}(a). The Majorana wavefunctions are significantly overlapped, indicating that the zero energy states are not MBSs.

\end{widetext}

\end{document}